\documentclass[lettersize,journal]{IEEEtran}
\usepackage{graphicx}
\usepackage{multirow}
\usepackage[flushleft]{threeparttable}
\usepackage{booktabs}
\usepackage{caption}
\usepackage{subcaption}
\usepackage[framemethod=tikz]{mdframed}
\usepackage{amssymb,amsmath,amsfonts,booktabs,times,float}
\usepackage{algorithm}
\usepackage{algpseudocode}
\usepackage{array}
\usepackage{textcomp}
\usepackage{amsmath}
\usepackage{url}
\usepackage{cite}
\usepackage{verbatim}
\usepackage{graphicx}
\usepackage{stfloats}
\usepackage[table]{xcolor}
\usepackage{colortbl}
\usepackage{xurl}
\usepackage{enumitem}
\usepackage[colorlinks=true, linkcolor=red]{hyperref}

\usepackage[]{review}
\setrevision{1}
\begin{document}
\setcounter{table}{0}

\title{LS-EEND: Long-Form Streaming End-to-End Neural Diarization with Online Attractor Extraction}

\author{Di Liang, Xiaofei Li
        % <-this % stops a space
\thanks{Di Liang is with Zhejiang University and also with Westlake University,
Hangzhou, China, e-mail: liangdi@westlake.edu.cn. Xiaofei Li is with the School of Engineering, Westlake University, Hangzhou 310030, China, and also with the Institute of Advanced Technology, Westlake Institute for Advanced Study, Hangzhou 310024, China. Corresponding author:  Xiaofei Li, e-mail: lixiaofei@westlake.edu.cn.}% <-this % stops a space
}

% {Shell \MakeLowercase{\textit{et al.}}: Frame-wise Streaming End-to-End Neural Diarization with linear-complexity inference}


\maketitle

\begin{abstract}
This work proposes a frame-wise online/streaming end-to-end neural diarization (EEND) method, which detects speaker activities in a frame-in-frame-out fashion. The proposed model mainly consists of a causal embedding encoder and an online attractor decoder. Speakers are modelled in the self-attention-based decoder along both the time and speaker dimensions, and frame-wise speaker attractors are automatically generated and updated for new speakers and existing speakers, respectively. Retention mechanism is employed and especially adapted for long-form diarization with a linear temporal complexity. A multi-step progressive training strategy is proposed for gradually learning from easy tasks to hard tasks in terms of the number of speakers and audio length. Finally, the proposed model (referred to as long-form streaming EEND, LS-EEND) is able to perform streaming diarization for a high (up to 8) and flexible number speakers and very long (say one hour) audio recordings. 
% A look-ahead mechanism is utilized to leverage some future frames for improving embedding quality.  Moreover, the Conformer encoder is adopted to enhance local interactions. Additionally, we explore a multi-step progressive training strategy to adapt to audio streams with varying number of speakers and lengths. 
Experiments on various simulated and real-world datasets show that: 1) when not using oracle speech activity information, the proposed model achieves new state-of-the-art online diarization error rate on all datasets, including CALLHOME (12.11\%), DIHARD II (27.58\%), DIHARD III (19.61\%), and AMI (20.76\%); 2) Due to the frame-in-frame-out processing fashion and the linear temporal complexity, the proposed model achieves several times lower real-time-factor than comparison online diarization models. Code is available on our github page \footnote{\url{https://github.com/Audio-WestlakeU/FS-EEND}}.
\end{abstract}


\section{Introduction}
\label{sec_1_intro}
% \addnote[speech science]{1}{\IEEEPARstart{V}{ariations} in vocal production organs, such as vocal tract and larynx, along with differences in speaking styles, including prosody and pauses, shape the uniqueness of each individual's voice, enabling speaker identification \cite{hanifa2021review}.} 
\IEEEPARstart{S}{peaker} diarization is a task of identifying speakers and their active time intervals within an audio recording, aiming to determine "who spoke when" in a multi-speaker scenario \cite{park2022review}. This technique is extensively utilized in real-world applications, including video conferences \cite{sawalatabad2021meeting}, medical systems \cite{finley18_interspeech}, and telephone speech analysis \cite{SERAFINI2023telephone}. Additionally, it serves as a crucial pre-processing step for many speech processing tasks. Leveraging the estimated utterance boundaries is essential for improving accuracy of automatic speech recognition (ASR) \cite{cornell2024joint} and speech translation \cite{yang2024st}. Moreover, jointly modeling speaker diarization and speech separation has been shown to be complementary to each other \cite{maiti2023ss, taherian2024ss}.

Conventional speaker diarization methods are developed based on a cascaded framework \cite{zhang2019fully}, consisting of four sequential stages: speech activity detection (SAD), speech segmentation, embedding extraction, and clustering. Although clustering-based methods can handle a variable number of speakers, they struggle with speech overlap, as each frame is assigned to only one speaker. Additionally, each module in the cascaded system is trained independently, which makes it difficult to optimize the whole diarization system. 

End-to-end approaches have been proposed to address these issues \cite{fujita2019end1, fujita2019end2, horiguchi2020end, Horiguchi2022taslp, Horiguchi2021end}. In \cite{fujita2019end1, fujita2019end2}, the diarization task is formulated as a multi-label classification problem and an end-to-end neural diarization (EEND) model is designed to compute frame-wise activity probabilities for each speaker. Permutation invariant training (PIT) loss is employed to cope with the speaker permutation ambiguity \cite{yu2017pit}. The fixed output dimension of EEND in \cite{fujita2019end1, fujita2019end2} limits the number of speakers it can handle. An extended EEND model with a LSTM-based encoder-decoder module is proposed to accommodate a flexible number of attractors, which serve as speaker centroids \cite{horiguchi2020end}. The attention mechanism is utilized to replace LSTM to better model the global context of embeddings \cite{rybicka2023attentionbased, chentaslp2024}, thereby enhancing diarization performance.

Offline diarization systems are effective for applications like meeting minutes. However, in some online scenarios, such as multi-party human-robot interaction or real-time subtitling of video conferences, the system needs to process audio streams and recognize/respond to the speaking person in real time. To meet these demands, several online/streaming diarization methods have been proposed \cite{coria2021overlap, zhang22_odyssey, yue22b_interspeech, kwon2023absolute, gruttadauria2024online, Fini2020supervised, han2021bw, xue2021online1, xue2021online2, horiguchi2022online}. In an online cascaded system, each module needs to operate in real time. Typically, sufficient audio frames are collected within a chunk (e.g., 2 s) for embedding extraction, and a short chunk shift (e.g., 0.5 s) is used to achieve low diarization latency \cite{coria2021overlap, zhang22_odyssey, yue22b_interspeech, kwon2023absolute}, then the system performs incremental clustering of extracted embeddings.
Along the time increases, re-clustering from scratch for each new segment becomes time-consuming. One solution is to cluster with centroids, which tracks global speaker label information \cite{kwon2023absolute, coria2021overlap}. Another approach is to select representative segments for local clustering, which requires addressing the permutation ambiguity problem \cite{zhang22_odyssey, yue22b_interspeech}. For end-to-end models, the offline EEND is adapted for online inference in \cite{xue2021online1, xue2021online2, horiguchi2022online}. A speaker tracing buffer (STB) is introduced to restore previous frames and diarization results. For each new chunk, the new frames are stacked with the buffered frames to perform local diarization. The new diarization results are then permuted according to the buffered diarization results to ensure consistent speaker order. Since all the buffered frames need to be re-fed to the diarization model for each chunk inference, STB-based methods require redundant computations. Additionally, a frame selection strategy needs to be designed carefully to ensure that the buffer is informative enough to solve the speaker permutation ambiguity. Recently, online diarization methods based on target-speaker tracking are proposed to perform chunk-wise inference by extracting target-speaker embeddings in real time \cite{wang2023endtoendonlinespeakerdiarization, cheng2024sequence}.

Different from the block- or chunk-wise methods, in this work, we propose a novel frame-wise streaming end-to-end neural diarization model, which processes audio streams in a frame-in-frame-out manner. The proposed model mainly consists of a causal embedding encoder and a non-autoregressive online attractor decoder. At each frame, the encoder extracts speaker embedding, and the decoder updates the attractor of each speaker, then diaraization results are obtained by taking the inner product between speaker embedding and attractors. The non-autoregressive decoder processes a two-dimensional sequence including the time dimension and speaker dimension, where each speaker inhabits one step of speaker sequence. 
The length of attractor/speaker sequence is set to a constant value, and the model can process flexible number of speakers that is not larger than this maximum value. In this work, we set the sequence length to a large value, i.e. 8, which is sufficiently high for lots of real applications. 
% \addnote[fixed point]{1}{Attractors are dynamically generated and updated over time, and are expected to converge to the centroid of embeddings corresponding to active speakers -- a process that is analogous to convergence toward fixed points.}

For embedding extracting and attractor updating, the encoder and decoder retrieve useful historical information through a self-attention mechanism, respectively. 
For long-form diarization (e.g. for one hour of audio recordings), the quadratic temporal complexity of self-attention brings significant computational challenge, thence we adapt the Retention mechanism \cite{retnet} into our model to achieve a linear temporal complexity and a constant computation cost as the time increases. Retention removes the softmax function when computing the self-attention values, so that the computation can be formulated in a recursive way and has a linear temporal complexity.

In addition, a look-ahead mechanism is implemented to leverage some future frames for making more confident decisions. Especially when a new speaker appears or a new utterance starts, there is no sufficient information for making reliable decisions, leveraging future information would be helpful. An embedding similarity loss is introduced to facilitate the training of encoder. 

Overall, the proposed model is designed to perform streaming end-to-end diarization for flexible number of speakers and varying audio lengths. However, directly training the model with a high number of speakers and long audio recordings is difficult. To resolve this problem, we propose a multi-step training strategy which progresses from easy to difficult tasks in terms of both the number of speakers and audio length. This strategy is effective for gradually learning more difficult tasks without sacrificing the performance of easy tasks. The proposed model is referred to as LS-EEND for its core characteristic of Long-form Streaming EEND. 
Experiments are conducted with multiple simulated and real-world datasets, on which the proposed model consistently outperforms other state-of-the-art (SOTA) online diarization methods, with several times lower computational cost.  

% Considering the significant challenges to computation time and memory as the length of audio streams increases, we adapt the Retention \cite{retnet} mechanism into our model to achieve linear-complexity inference, resulting in the LS-EEND system. 

% file is intended to serve as a ``sample article file''
% for IEEE journal papers produced under \LaTeX\ using
% IEEEtran.cls version 1.8b and later. The most common elements are covered in the simplified and updated instructions in ``New\_IEEEtran\_how-to.pdf''. For less common elements you can refer back to the original ``IEEEtran\_HOWTO.pdf''. It is assumed that the reader has a basic working knowledge of \LaTeX. Those who are new to \LaTeX \ are encouraged to read Tobias Oetiker's ``The Not So Short Introduction to \LaTeX ,'' available at: \url{http://tug.ctan.org/info/lshort/english/lshort.pdf} which provides an overview of working with \LaTeX.

This paper is an extension of our previous conference paper, i.e. FS-EEND \cite{liang2024framewise}. The new contributions of this extension are as follows:
\begin{enumerate}
\item{We adapt Retention to replace the masked self-attention used in \cite{liang2024framewise}. This way, the model has a linear temporal complexity, and thus can be used for long-form diarization.}
\item{We enhance the Transformer encoder used in \cite{liang2024framewise} with Conformer encoder, and validate its superiority on multiple diarization datasets.}
\item{We propose a new multi-step progressive training strategy to enable the model to process a high number of speakers and very long audio recordings.}
\item{We conduct through experiments on multiple simulated and real-world datasets, and compare with more SOTA online diarization methods, to fully evaluate the proposed model under diverse conditions.}
% \item{We also investigate the impact of varying training chunk lengths on real datasets.}
\end{enumerate}

The organization of this paper is as follows. Section II reviews and compares with related works. Section III describes the proposed LS-EEND system in detail. Section IV and V presents the experimental setup and results, respectively. Section VI concludes the paper.

\section{Related Works}

% \subsection{Offline Diarization}
% \subsubsection{Cascaded System} 
Speaker diarization is essentially a speaker-oriented frame clustering problem. Cascaded diarization system \cite{garcia2017segment} separately performs frame/segment-wise speaker embedding extraction and speaker embedding clustering, then speaker diarization is completed based on the cluster assignment of speaker embeddings. Pre-trained speaker embedding networks \cite{snyder2018x, mary2022svector} are normally used, which provide unchanging speaker embeddings, namely one speaker is always represented with the same embedding when he/she presents in different contexts. Clustering can be performed with traditional methods, such as hierarchical clustering \cite{bouguettaya2015efficient} and spectral clustering \cite{huang2019ultra}. For online diarization, one core difficulty is to make the clustering step online through online/incremental clustering algorithms \cite{coria2021overlap,zhang22_odyssey,yue22b_interspeech, kwon2023absolute}, where clustering centroids or representative frames should be well organized and memorized to keep the speaker consistency over a long time period. 
A recurrent neural network (RNN)-based online clustering method is proposed in UIS-RNN \cite{zhang2019fully}, where each speaker is modeled by an instance of RNN, and RNN instances can be generated for flexible number of speakers. 

% Cascaded diarization system  is composed of several sub-modules. Firstly, a speech activity detector filters out non-speech segments from the audio stream \cite{gelly2018sad}. The remaining speech region is then segmented \cite{garcia2017segment} into shorter segments. Subsequently, speaker embeddings are extracted for each segment using a pre-trained speaker embedding extractor \cite{dehak2010front}. These embeddings are then clustered based on a suitable similarity measurement, such as cosine similarity or PLDA \cite{kenny2013plda}. Typical clustering methods employed for diarization include K-means \cite{dimitriadis2017developing}, hierarchical clustering \cite{bouguettaya2015efficient}, and spectral clustering \cite{huang2019ultra}, etc. Optionally, a resegmentation post-processing module is used to refine the clustering results. Many efforts are devoted to implementing the embedding extraction module with a well-designed deep neural network to enhance the quality of speaker embeddings \cite{snyder17_emb, snyder2018x, mary2022svector}.

EEND methods integrates speaker embedding extraction and clustering  into a single network \cite{fujita2019end1, fujita2019end2}. To handle flexible number of speakers, Horiguchi et al. \cite{horiguchi2020end,Horiguchi2022taslp} propose the EEND-EDA system, where a flexible number of attractors are generated by an encoder-decoder-based attractor extractor. Different from the pre-trained unchanging speaker embeddings \cite{snyder2018x, mary2022svector},  EEND methods generate changing/local speaker embeddings, namely one speaker is represented with different embeddings when he/she presents in different contexts. Although the  extractor can output attractor for unlimited number of speakers, the performance will largely drop when test speakers are more than training speakers. To really handle unlimited number of speakers, \cite{Horiguchi2021ASRU,horiguchi2022online} proposes to further transform the local attractors obtained within signal blocks to global attractors, then the global attractors can be clustered with an unsupervised clustering method. Note that the number of speakers within one short block is limited, while the number of global speakers can be unlimited.  

Based on the offline EEND methods, several block-wise online EEND methods have been proposed \cite{han2021bw, xue2021online2,horiguchi2022online}, which need to solve the speaker permutation ambiguity between blocks. The BW-EEND-EDA method \cite{han2021bw} transfers the hidden state of attractor extractor across blocks to solve the speaker permutation ambiguity. Speaker tracing buffer (STB) is proposed in \cite{xue2021online2} to store previous feature frames and diarization results, which is then used together with a new feature block for diarization inference, and  the speaker permutation ambiguity can be resolved by aligning the diarization results of buffering frames. In \cite{horiguchi2022online}, STB is further improved by applying block-wise buffer updating. 

At a high level, the proposed LS-EEND model follows the principle of EEND-EDA, namely using an attractor extractor for handling flexible number speakers, and training the speaker embedding encoder and the attractor extractor together. However, the proposed method is prominently different from the block-wise online EEND methods \cite{han2021bw, xue2021online2,horiguchi2022online} developed based on EEND-EDA, as the proposed method leverages one single model to frame-wisely update speaker attractors and maintain speaker consistency along the entire audio stream. Actually, the proposed attractor decoder shares a similar spirit with UIS-RNN \cite{zhang2019fully} in the sense that both models accommodate each speaker in a network, and the historical information of the speaker are stored in and can be retrieved from the network's hidden states, which is crucial for accurately maintaining speaker consistency over long periods of time. However, the implementation of the two models are totally different.

\section{The Proposed Frame-wise Streaming End-to-End Neural Diarization System}
This section provides a detailed description of the proposed LS-EEND system. Firstly, we briefly introduce the speaker order of appearance to accommodate streaming applications. Next, we present the proposed end-to-end causal diarization model. Finally, we outline the training strategy and inference process.

\subsection{Label Permutation based on Speaker Appearance Order}
In offline scenarios, frame-level speaker embeddings are clustered to determine speaker activity timestamps, where the order of speakers is arbitrary. The diarization model can be trained using the PIT loss \cite{yu2017pit}, similar to techniques employed in the source separation field \cite{quan2022pit}. However, in streaming scenarios, speakers appear sequentially. The speakers' order should thus align with their appearance order.

\begin{figure}[!t]
\centering
\includegraphics[width=0.9\linewidth]{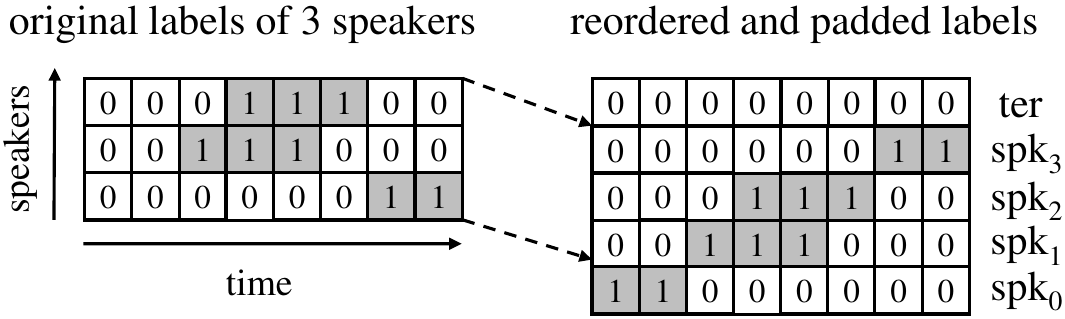}
\caption{An example of label permutation according to the speaker appearance order. $\text{spk}_{\text{0}}$ represents an extra non-speech speaker. `ter' stands for the termination of active speakers.}
\label{fig:transf_label}
\end{figure}

Fig. \ref{fig:transf_label} illustrates an example of diarization label permutation according to speaker appearance order. The label value 1 indicates active speech, while 0 inactive speech. An additional speaker is introduced for non-speech frames, denoted as spk$_0$.   
The introduction of non-speech speaker trains the network to explicitly integrate SAD as an auxiliary task, which has been shown to be able to reinforce  speaker diarization \cite{2021takashima_multitask}. The actual speakers are sequentially labeled from spk$_\text{1}$ to spk$_s$ following their appearance order. To determine the number of speakers, we append an explicit speaker termination marker (zero label) above the active speaker labels, serving as a speaker counter. This adjustment increases the number of speaker tracks from the number of active speakers $S$ to $S+2$.

% Fig. 1 is an example of a floating figure using the graphicx package.
%  Note that $\backslash${\tt{label}} must occur AFTER (or within) $\backslash${\tt{caption}}.
%  For figures, $\backslash${\tt{caption}} should occur after the $\backslash${\tt{includegraphics}}.

\subsection{Model Architecture}
\label{sec_3b_model_arch}
\begin{figure*}[!t]
\centering
\includegraphics[width=0.85\linewidth]{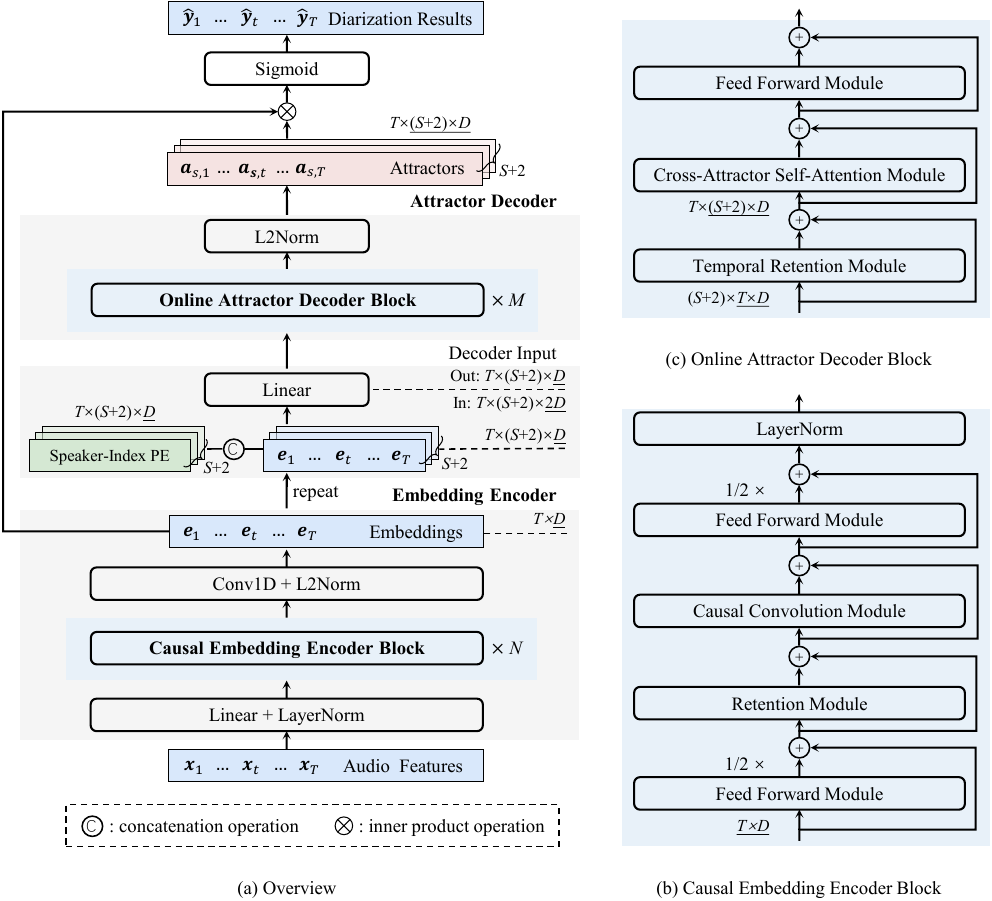}
\caption{Network architecture of the proposed LS-EEND model. `$S$+2' is the number of speakers corresponding to the augmented speaker label in Fig.~\ref{fig:transf_label}. The data dimensions are presented in the form of ``batch dimension $\times$ \underline{dimension of one sample in batch}".}
\label{fig:model_archi}
\end{figure*}

The architecture of the proposed causal neural diarization model is shown in Fig. \ref{fig:model_archi}, which mainly consists of a speaker embedding encoder and an attractor decoder. The input LogMel feature sequence is denoted as $\boldsymbol{X} = (\boldsymbol{x}_1, \ldots, \boldsymbol{x}_t, \ldots, \boldsymbol{x}_{T})$ with $ \boldsymbol{x}_t \in \mathbb{R} ^ {F}$, where $T, F$ stands for the number of frames and the feature dimension, respectively. The input vectors are normalized with the cumulative mean vector as $\boldsymbol{x}_{t}-\boldsymbol{\mu}_{t}$, where $\boldsymbol{\mu}_{t}=\frac{t-1}{t} \boldsymbol{\mu}_{t-1}+\frac{1}{t} \boldsymbol{x}_t$.

\subsubsection{Causal Embedding Encoder}
The causal embedding encoder takes as input the feature sequence $\boldsymbol{X}$ and extracts frame-wise speaker embeddings $\boldsymbol{E}=(\boldsymbol{e}_1, \ldots, \boldsymbol{e}_t, \ldots\boldsymbol{e}_{T})$, with $\boldsymbol{e}_t \in \mathbb{R} ^ {D}$ where ${D}$ denotes embedding dimension. As shown in Fig.~\ref{fig:model_archi}(b), the proposed encoder block employs the architecture of Conformer \cite{gulati20_interspeech}, which has been validated effectiveness for multiple speech processing tasks \cite{borsos2023conformertts, chen2022conformerss}. The multi-head Retention module is followed by a convolution module then a feed-forward module. Retention and convolution are expected to capture global and local dependencies, respectively. It should be noted that the convolution module in our implementation of Conformer is modified to be causal. $N$ encoder blocks are cascaded.

In most of speaker embedding networks, self-attention-based encoders are normally adopted for learning long-term dependency. As for streaming diarization, the system needs to continuously process significantly long audio streams (e.g. a meeting of hours) in an online manner. The masked self-attention mechanism can realize online processing by preventing the use of future information, such as in our previous work \cite{liang2024framewise}, but it faces a huge computational challenge due to the quadratic temporal complexity of self-attention. Linear temporal complexity can be achieved by truncating the masked self-attention with a constant truncation length, which however constrains the retrieval of full  historical information. In this work, we adapt the Retention mechanism \cite{retnet}, which has a linear temporal complexity and retains the capability of full history retrieval. As a successor to the (masked and causal) self-attention mechanism, Retention removes the softmax function on the attention values, so that the computation of attention can be formulated in a recursive way and has a linear complexity w.r.t time steps. Similar to self-attention, for a given input sequence $X\in \mathbb{R} ^ {T\times D}$, Retention also computes the key, query and value vectors, denoted as $K, Q, V \in \mathbb{R}^{T \times D}$, with trainable matrices $W_K, W_Q, W_V \in \mathbb{R}^{D \times D}$, then aggregates the value vectors. However, the specific computations are somehow different. Specifically, Retention can be formulated in a parallel way as:
\begin{align}
     &\text{Retention}(X)=(QK^{\top}\odot \Gamma)V \nonumber \\
    Q=(X&W_Q)\odot \Theta, K=(XW_K)\odot \bar{\Theta}, V=XW_V,
\label{eq:retention-parallel}
\end{align}
where $^{\top}$ and $\odot$ denote matrix/vector transpose and element-wise product, respectively. $\Theta, \bar{\Theta} \in \mathbb{C}^{T \times D/2}$ ($\Theta_{t}=e^{it\theta}$, $\bar{\Theta}_{t}=e^{-it\theta}$) and $\Gamma \in \mathbb{R}^{T \times T}$  together provide the so-called xPos \cite{SDP} as a relative rotary position embedding, where $i=\sqrt{-1}$ denotes the imaginary unit, $\theta \in \mathbb{R}^{1\times D/2}$ is a constant vector. The relative positional embedding terms, i.e. $\Theta$ and $\bar{\Theta}$ are removed in this work as positional embedding has no significant impact on the model's performance, similar to the findings in the previous neural diarization methods \cite{Horiguchi2022taslp, chentaslp2024}.
% The complex multiplication in Eq.~(\ref{eq:retention-parallel}) corresponds to vector rotation and is implemented as rotary position embedding \cite{SU2024roformer}.} %$\theta$ $\gamma$ are two constant real numbers
$\Gamma$ ($\Gamma_{t_1t_2}=\gamma^{t_1-t_2}$, \text{for} $t_2\le t_1$; 0 otherwise) is a lower triangular mask matrix, which combines causal masking (similar to masked self-attention \cite{NICOLSON202080}) and exponential decay, where $\gamma \in \mathbb{R}$ is a constant exponential decay factor.
It can be seen that, different from the self-attention mechanism, the softmax function is not applied to the attention scores of $QK^{\top}\odot \Gamma$. By doing this, Retention can be re-formulated in a recurrent way as:
\begin{align}
     &\text{Retention}(X_t)=Q_tS_t \nonumber \\
   & S_t=\gamma S_{t-1}+K_t^{\top}V_t,
\label{eq:retention-recurrent}
\end{align}
where $K_t, Q_t, V_t \in \mathbb{R}^{1\times D}$ are the key, query and value vectors at frame $t$, $S_t \in \mathbb{R}^{D \times D}$ is the state matrix which is recursively updated along time. The parallel representation Eq.~(\ref{eq:retention-parallel}) is suitable for efficient training, while the recurrent representation Eq.~(\ref{eq:retention-recurrent}) has a linear temporal complexity and can be used for inference. In between, \cite{retnet} also provides a chuckwise recurrent representation, which facilitates the training with long sequences. Within each chunk, computation follows the parallel representation as shown in Eq.~(\ref{eq:retention-parallel}). While state recurrence is performed across chunks using the recurrent representation in Eq.~(\ref{eq:retention-recurrent}). Specifically, let $B$ and $i$ denote the length and index of chunk, respectively. The input sequence, key, query and value of the $i$-th (non-overlapped) chunk are denoted as $X_{[i]}, K_{[i]}, Q_{[i]}, V_{[i]} \in \mathbb{R}^{B \times D}$, respectively. The chunk-wise state matrix $R_{i}\in \mathbb{R}^{D \times D}$ can be recursively computed as: 
\begin{align}
R_{i} = K_{[i]}^{\top}(V_{[i]} \odot \zeta)+\gamma^{B}R_{i-1}, \ \zeta_{mn}=\gamma^{B-m-1}, 
\end{align}
and then the chunk-wise Retention can be computed as
\begin{align}
\text{Retention}(X_{[i]})=&\underbrace{(Q_{[i]}K_{[i]}^{\top} \odot \Gamma )V_{[i]}}_{\text{Inner-chunk}}+ 
\underbrace{(Q_{[i]}R_{i-1}) \odot \xi}_{\text{Cross-chunk}}, \nonumber \\ &\xi_{mn}=\gamma^{m+1},
\label{eq:retention-chunkwise}
\end{align}
where $\Gamma \in \mathbb{R}^{B \times B}$ and $\gamma$ are similarly defined as in Eq.~(\ref{eq:retention-parallel}).    
This design allows Retention to maintain linear memory complexity for processing long sequences.

% supports three computation paradigms: parallel, recurrent, and chunk-wise recurrent. The parallel paradigm allows for fast training. The recurrent representation enables $O(n)$ inference regarding memory and computation cost, where $n$ denotes sequence length. 
% As a hybrid of parallel and recurrent representations, the chunk-wise paradigm performs parallel computation within chunk and sate recurrence across chunk, accelerating training for long sequences.
Removing softmax could degrade the numerical scale stability. To compensate for this, three extra normalization factors are applied to the attention scores of $QK^{\top}\odot \Gamma$ in Eq.~(\ref{eq:retention-parallel}) and a group normalization layer is added right after the retention layer, please refer to \cite{retnet} for more details.

Retention also uses multiple heads, and multi-scale retention assigns different $\gamma$ for each head as
\begin{align}
    \gamma = 1 - 2 ^ {-5-\text{arange}(0, h)} \in \mathbb{R} ^ {h} 
\label{eq:gama}
\end{align}
where $h$ denotes the number of Retention heads. $\gamma$ controls the forgetting rate of past information. In addition, after the multi-head retention and the group normalization layer, an extra swish gate \cite{ramachandran2017swish} is applied to increase the non-linearity of retention layers. 

% To compensate for this, a GroupNorm strategy, three normalization factors, and a swish gate \cite{ramachandran2017swish} are implemented to achieve scale invariance and increase the nonlinearity in \cite{retnet}. Formally, the output of a Retention layer is defined as:
% \begin{align}
%     \gamma = 1 - 2 ^ {-5-\text{arange}(0, h)} \in \mathbb{R} ^ {h} 
% \end{align}
%     \text{head}_{i}&=\text{Retention}(X,\gamma_{i}) \nonumber \\
%     Y &= \text{GroupNorm}_{h}(\text{Concat}(\text{head}_{1}, \ldots, \text{head}_{h})) \nonumber \\
%     \text{MSR}(X) &= (\text{swish}(XW_{G})\odot Y)W_{O},
% \label{eq:gama}
% \end{align}
% % The vanilla RetNet introduces exponential decay along relative distance through parameter $\gamma$ for information forgetting, defined as
% % \begin{equation}
% %     \gamma = 1 - 2 ^ {-5-\text{arange}(0, h)} \in \mathbb{R} ^ {h}
% % \label{eq:gama}
% % \end{equation}
% where $h$ denotes the number of Retention heads. $\gamma$ controls the forgetting rate of past information. $W_G$, $W_O \color{blue} \in \mathbb{R}^{D \times D}$ are trainable matrices.}  

Speaker diarization learns voiceprint of speakers, which are time-invariant in a very long time period, e.g. in the entire audio stream to be processed. In the proposed model, to keep the consistency of speaker embedding along the entire audio stream,  we set $\gamma=1$ to not have information decay.
% In addition, the relative positional embedding terms, i.e. $\Theta$ and $\bar{\Theta}$ in Eq.~(\ref{eq:retention-parallel}), are removed as positional embedding has no significant impact on the model's performance, similar to the findings in the previous neural diarization methods \cite{Horiguchi2022taslp, chentaslp2024}. 
% While in the diarization task, audio frames are consistently important regardless of their occurrence time. Frames appearing earlier do not diminish in importance over time for subsequent speaker recognition. For example, in a one-hour meeting, the host delivers opening remarks in the first five minutes and closing remarks in the last five minutes. When the system performs diarization for the last five minutes, it needs to attend to the audio streams of the first minutes. 

\subsubsection{Convolutional Look-ahead}
For most online/streaming applications, a short response latency is acceptable. Leveraging some future frames can significantly improve the quality of speaker embeddings. Especially, the birth of a new speaker is subtle and difficult to detect, when the speaker appears in only a few frames. Therefore, we implement a look-ahead mechanism to utilize a few future frames and make more confident decisions, by adopting 1-dimensional convolution along the time dimension. The latency, defined as the number of utilized future frames, is determined by the convolution kernel and padding size. 

Finally, the embeddings are normalized by L2-norm, which is widely used in representation learning \cite{li2024ssl}, to enhance the stability of embeddings. 

\subsubsection{Online Attractor Decoder}
In contrast to the block-wise online diarization methods \cite{han2021bw, horiguchi2022online}, which generate block-wise attractors to perform diarization for the entire block, we design an online attractor decoder to extract attractors frame by frame. This approach eliminates the need for buffer design, allowing the network to automatically select and utilize previous information, such as previous representative acoustic features and attractors of appeared speakers. The decoder takes as input the embeddings, and is designed to immediately detect new speakers when they appear, and to update  the attractors of existing speakers as more speaker embeddings are collected over time. Unlike sequence generation tasks such as speech recognition or machine translation, the multiple attractors in diarization task do not follow a certain ordered semantic dependency, making parallel processing more effective. Therefore, we propose a non-autoregressive attractor decoder to simultaneously generate multiple attractors. The architecture of the proposed attractor decoder is depicted in Fig. \ref{fig:model_archi}(a) and (c), including decoder input, decoder block, and decoder output. The decoder data has two sequence dimensions: the time-frame dimension $T$ and the attractor dimension $S+2$.

\textbf{Decoder Input.} At each frame, the speaker embedding is fed into the decoder to generate/update the corresponding attractors. All the $S+2$ attractors share the same repeated embedding as their input source, and the decoder determines whether the embedding is used for attractor updating. To distinguish the input to multiple attractors from each other, we first repeat the speaker embedding $S+2$ times, which are then concatenated (in the feature dimension) with a positional embedding (PE) \cite{vaswani2017attention} along the speaker/attractor index (speaker-index PE): 
\begin{equation}
\begin{split}
\text{Speaker-Index PE} (s, 2d) &= sin(s/10000^{2d/D)}, \\
\text{Speaker-Index PE} (s, 2d+1) &= cos(s/10000^{2d/D}),
\end{split}
\end{equation}
where $s$ indexes the speakers/attractors, and $d$ indexes the feature dimension. Note that the same speaker-index PE is utilized for all time frames. 
% The difference in input to multiple attractors at each time frame arises from the varying indices of speakers/attractors.
A linear layer is then applied to restore the feature dimension $D$. Thus, the input dimension of the attractor decoder is ${T \times (S+2) \times D}$.

\textbf{Decoder Architecture.} To derive the attractor $\boldsymbol{a}_{s,t}$ of the $s$-th speaker at frame $t$, besides the speaker embedding extracted at frame $t$, two more information sources need to be considered:
\begin{itemize}[leftmargin=*]
\item The attractor of previous frames for the same speaker,  i.e., $\boldsymbol{a}_{s,t^{\prime}}, t'< t$, which helps maintain the attractor's consistency of a specific speaker over time. To refer to the extracted attractor of previous frames,  recursive computation is normally required, which however causes implementation difficulties of parallel computation during training. Instead, we design a non-recursive scheme by applying a temporal Retention module in the hidden space of attractors, which is also proved to be effective for retrieving the attractor information of previous frames. Specifically, let $H \in \mathbb{R} ^ {T \times (S+2) \times D}$ denote the hidden units of attractors. The temporal Retention module processes the hidden sequence of each speaker/attractor independently, and the attractor axis is taken into the batch dimension:   
\begin{equation}
\text{Temporal Retention}(H[:, s, :]). 
\end{equation}
% The consistency of a speaker's attractor over time is modeled by the temporal formulation of Retention.
 
For the same reason as stated above, in temporal Retention, $\gamma$ is set to 1 and the positional embedding terms are removed.
% The Retention structure ensures causality and $O(n)$ inference. 
 \item The attractor of other speakers at frame $t$, i.e., $\boldsymbol{a}_{s^{\prime},t}, s'\neq s$. By leveraging information from other attractors, the distance between attractors can be enlarged to better distinguish different speakers. To this end, we develop a cross-attractor self-attention module which processes each time frame independently, following the standard multi-head self-attention mechanism \cite{vaswani2017attention}:
\begin{equation}
\text{Cross-Attractor Self-attention}(H[t, :, :]).
\end{equation}
%where $K, Q, V \in \mathbb{R}^{(S+2) \times D}$ are the key, query, and value matrices linearly projected from $\boldsymbol{A}_{t} \in \mathbb{R}^{(S+2) \times D}$, respectively. The cross-attractor self-attention module models the distinction between different attractors.

\end{itemize}
After, a feed-forward module is employed. Each of the three modules incorporates a residual connection and layer normalization. $M$ decoder blocks are cascaded.

\textbf{Decoder Output.} The decoder output vectors are then L2-norm normalized to have unit length, serving as the final attractors $\boldsymbol{A}_{t} = (\boldsymbol{a}_{0,t},\boldsymbol{a}_{1,t},\ldots,\boldsymbol{a}_{S,t},\boldsymbol{a}_{\text{ter},t})$. Speaker activities are calculated by taking the inner product between the frame-wise attractors and the speaker embedding $\boldsymbol{e}_t$ as: 
\begin{equation}
\hat{\boldsymbol{y}}_{t} = \sigma ( \boldsymbol{A}_t^{\top} \boldsymbol{e}_t ) \in (0, 1)^{S+2},
\end{equation}
where $\sigma(\cdot)$ denotes the element-wise sigmoid function. Here, $\hat{\boldsymbol{y}}_{t}=(\hat{y}_{0,t},\hat{y}_{1,t}\ldots,\hat{y}_{S,t},\hat{y}_{\text{ter},t})$ denotes the posterior probabilities of non-speech, $S$ active speakers, and termination of active speakers. 

\subsection{Training Loss}
\label{sec3c_mod_train}
The diarization loss is defined as the binary cross entropy (BCE) between the estimated posterior speech activities $\hat{\boldsymbol{Y}}$ and the ground truth $\boldsymbol{Y}$ (as shown in the right table of Fig.\ref{fig:transf_label})
\begin{equation}
\mathcal{L}_d=\frac{1}{T(S+2)} \sum^{T}_{t=1} \sum_{s \in \mathbb{S}} [-y_{s,t} \log \hat{y}_{s,t} - (1-y_{s,t}) \log(1 - \hat{y}_{s,t})],
\label{eq:dia_loss}
\end{equation}
where $\mathbb{S} =\{0, 1,\ldots, S, {\text{ter}}\}$. 
The diarization loss trains the network to: i) maintain the consistency of speaker embedding (and also its corresponding attractor) extracted at different frames for the same speaker, and disentangle out speech content information; ii) better discriminate speaker embeddings (and also the corresponding attractors) for different speakers, by leveraging such as their different vocal tract attributes (e.g., length and shape), vocal fold properties (e.g., fundamental frequency F0), and speaking styles (e.g., prosody including rhythm and intensity)\cite{wang2024overview, Douglas2024review}; and iii) to be able to handle overlapped speakers in terms of both speaker embedding learning and attractor learning, by performing kind of multi-class classification.

Note that both the embeddings and attractors are estimated within each recording and are only used for speaker diarization of that recording. Embeddings (and also attractors) could be different for the same speaker when appearing in different recordings. A similar observation is discussed in \cite{Horiguchi2022taslp}.

\textbf{Embedding similarity loss}.
The clustering effect of the frame-wise speaker embeddings has a salient impact on diarization performance. In addition to the indirect training of speaker embeddings made by the diarization loss. To regulate the distribution of these embeddings, we introduce an extra direct embedding similarity loss. The cosine similarity between each pair of embeddings ($\langle \boldsymbol{e}_j, \boldsymbol{e}_k \rangle, j, k \in \{1, \ldots, T\}$) is constrained to be consistent with the cosine similarity between their corresponding diarization labels ($\langle \boldsymbol{y}_j, \boldsymbol{y}_k \rangle, j, k \in \{1, \ldots, T\}$). The mean square error between the two cosine similarities is minimized as
\begin{align}
  \mathcal{L}_e = \frac{1}{T(T-1)/2} \sum_{j=1}^{T} \sum_{k=j+1}^{T} (\langle \boldsymbol{e}_j, \boldsymbol{e}_k \rangle - \langle \boldsymbol{y}_j, \boldsymbol{y}_k \rangle)^{2}, \label{eq: l_e}
\end{align}
where $\langle \cdot, \cdot \rangle$ denotes cosine similarity. It should be noted that the embedding similarity loss is suitable for various situations, including two frames from the same, different, and overlapped speakers. The higher the overlap between two frames, the smaller the angle between their corresponding embedding vectors should be.

The total training loss is defined as the sum of diarization loss and embedding similarity loss
\begin{equation}
\mathcal{L} = \mathcal{L}_d+\mathcal{L}_e.
\end{equation}

\subsection{Training Strategy}
\label{sec_train_strategy}

The challenges of streaming diarization lie in two aspects: 1) processing long audio streams, where maintaining the consistency of speaker identities in long recordings is particularly difficult; 2) handling a high and flexible number of speakers, where increased  speakers leads to higher task complexity. In real-world scenarios, more speakers often correlate with longer speech, further complicating the processing. 
% An optimized single model is expected to accommodate a flexible number of speakers and perform effectively across scenarios with varying audio stream lengths. 
STB-based online diarization methods \cite{horiguchi2022online, xue2021online2} perform block-wise (100 s) processing, which avoids the challenges associated with variable-length inference but needs to resolve the speaker permutation ambiguity between chunks.

In this work, we aim to leverage one single end-to-end model performing streaming diarization frame by frame, which can directly handle a high and flexible number of speakers and varying audio lengths, without requiring additional design or extensions during inference. It is crucial to develop an effective training strategy that allows the model better learning such challenging tasks. A trivial approach would be to train directly with a large number of speakers (up to 8 speakers in this work) and long audio recordings (at the hour-level in this work). However, this faces two difficulties: 1) directly learning such a complex task, the model may converge to a sub-optimal point, which will be verified in Section \ref{sec5c1_ami}; 2) the training cost can be excessively high. Although the proposed model has a linear complexity w.r.t time steps, processing very long audio recordings is still very time-consuming. 
% due to the need for a large number of training samples to achieve convergence. 
To address these challenges, we propose a multi-step training strategy that progresses from easy to difficult tasks in terms of both the number of speakers and audio length. Specifically, first we progressively train the model with linearly increased number of speakers. At this stage, the audio length is kept constant and relatively short (say 100 s), and a large amount of simulated training data are used to fully establish the model's foundational capability for handling a high number speakers.
% short speech during the pre-training stage, we linearly increase the number of speakers on simulated data to establish the model's foundational capability for handling multi-speaker short speech. 
Then, using a relatively small amount of real-world data, we progressively fine-tune the model with doubly increased audio length to enhance the model's ability for processing long recordings. This strategy gradually learns more difficult tasks and hopefully without sacrificing the performance of easy tasks. Detailed settings for this strategy will be provided in Section \ref{sec4c_train_setup}.

\subsection{Inference}
\label{sec_infer}
% The block-wise diarization methods \cite{xue2021online2, horiguchi2022online} divert an offline model for online inference. They introduce STB to track speaker permutation information across blocks. All frames in the buffer need to be re-fed to the diarization network for each chunk inference. In contrast, 
At inference, the proposed model processes audio streams frame by frame, with the recurrent paradigm of Retention and a linear computational complexity w.r.t frames. 
% The  the overall system complexity is $O(n)$. performs the diarization processing once per frame. During inference stage, 
The maximum number of speakers $S$ is set to the maximum number appeared in the training set. Regardless of the actual number of speakers, the network always outputs activity probabilities for $S+2$ attractors. Since the first attractor represents non-speech, speech activity are evaluated starting from the second attractor. When the actual number of speakers is smaller than the preset $S$, it can be automatically detected by the speaker termination marker.
% As a result, $S+2$ attractors are generated by the decoder. Since the first attractor represents non-speech, posterior probabilities of speech activity are computed starting from the second attractor. 

\IEEEpubidadjcol

\section{Experimental Setup}
\subsection{Datasets}

\subsubsection{Simulated Mixtures}
\label{sec_sim_data}
We follow the recipe presented in \cite{horiguchi2022online} to simulate speech mixtures using Switchboard-2 (Phase I \& II \& III), Switchboard Cellular (Part 1 \& 2), and 2004-2008 NIST Speaker Recognition Evaluation (SRE) corpora. 
% \addnote[multi utts]{1}{Each mixture typically includes \emph{multiple, distinct} utterances per speaker, encouraging the model to disentangle the extracted embeddings and attractors from the specific speech content.} 
The corpus contains multilingual utterances in over twenty languages, the randomly generated mixtures include some multilingual scenarios. Background Noise from MUSAN corpus \cite{snyder2015musan} and Simulated Room Impulse Response \cite{ko2017study} are applied to the synthesized mixtures. The sampling rate is 8 kHz. The detailed statistical information of the simulated dataset are provided in Table \ref{tab:simdata}. For each of the 1$\sim$8 speakers cases, 100,000 mixtures are generated for training. The overlap ratios between speakers are set quite high, i.e. about 29\% to 35\%. Audio duration is accordingly increased when more speakers are involved. Besides, for each of the 1$\sim$8 speakers cases, another 500 mixtures are generated for test.

\begin{table}[t]
\renewcommand\arraystretch{1.15}
    \caption{\textbf{Simulated Mixtures for Training.} Ovl and Avg\_dur stand for overlap ratio and average duration, respectively. $\beta$  controls the intervals of multiple utterances of the same speaker. The larger $\beta$, the larger the intervals. }
    % \footnotesize
    \label{tab:simdata}
    \centering
            \resizebox{\columnwidth}{!}{
            \begin{tabular}{cccccccc}
                \toprule
                  & \#Spk & \#Mixtures & $\beta$ & Ovl (\%) & Avg\_dur (mins) \\
                \midrule
            \multirow{1}{*}{Sim1spk}   & 1 & 100,000 & 2 & 0.0    & 1.29   \\
                                    % ~ & Test  & 1 & 500     & 2 & 0.0    & 1.34   \\
            \midrule
            \multirow{1}{*}{Sim2spk}   & 2 & 100,000 & 2 &  34.7  & 1.49   \\
                                    % ~ & Test  & 2 & 500     & 2 &  35.2  & 1.50 \\
            \midrule
            \multirow{1}{*}{Sim3spk}   & 3 & 100,000 & 5 &  34.7  & 2.53  \\
                                    % ~ & Test  & 3 & 500     & 5 &  35.2  & 2.53  \\  
            \midrule
            \multirow{1}{*}{Sim4spk}   & 4 & 100,000 & 9 &  32.0  & 3.99   \\
                                    % ~ & Test  & 4 & 500     & 9 &  32.5  & 4.00   \\
            \midrule
            \multirow{1}{*}{Sim5spk}   & 5 & 100,000 & 13 &  30.8  & 5.52   \\
                                    % ~ & Test  & 5 & 500     & 13 &  30.8  & 5.49 \\
            \midrule
            \multirow{1}{*}{Sim6spk}   & 6 & 100,000 & 17 &  30.0  & 7.09   \\
                                    % ~ & Test  & 6 & 500     & 17 &  30.4  & 7.11 \\
            \midrule
            \multirow{1}{*}{Sim7spk}   & 7 & 100,000 & 21 &  29.6  & 8.73   \\
                                    % ~ & Test  & 7 & 500     & 21 &  29.8  & 8.68 \\
            \midrule
            \multirow{1}{*}{Sim8spk}   & 8 & 100,000 & 25 &  29.3  & 10.38   \\
                                    % ~ & Test  & 8 & 500     & 25 &  29.8  & 10.40 \\    
                \bottomrule
            \end{tabular}
    }
    % \vspace{1em}
\end{table}

\subsubsection{Real-world Recordings} We also evaluate the performance of the proposed method on various real-world datasets, including CALLHOME \cite{martin2001nist} \footnote{\url{https://catalog.ldc.upenn.edu/LDC2001S97}}, DIHARD II \cite{ryant2019seconddiharddiarizationchallenge}  \footnote{\url{https://catalog.ldc.upenn.edu/LDC2021S10} (and \href{https://catalog.ldc.upenn.edu/LDC2021S11}{LDC2021S11}, 
\href{https://catalog.ldc.upenn.edu/LDC2022S06}{LDC2022S06},  \href{https://catalog.ldc.upenn.edu/LDC2022S07}{LDC2022S07})}, DIHARD III \cite{ryant21dh3_interspeech} \footnote{\url{https://catalog.ldc.upenn.edu/LDC2022S12} (and \href{https://catalog.ldc.upenn.edu/LDC2022S14}{LDC2022S14})}, and AMI head set mix recordings \cite{ami} \footnote{\url{https://groups.inf.ed.ac.uk/ami/corpus}}. The detailed statistical information of these datasets are shown in Table \ref{tab:realdata}. The CALLHOME data is split into Part 1 and Part 2 according to the Kaldi recipe \footnote{\url{https://github.com/kaldi-asr/kaldi/tree/master/egs/callhome_diarization/v2}}, which are used for model adaptation and evaluation, respectively. For the DIHARD II and DIHARD III datasets, model adaptation and evaluation are conducted on the Dev set and Test set, respectively. The AMI headset mix recordings are divided into three subsets: Train, Dev, and Test. We perform model adaptation using the Train set and evaluation on the Dev and Test sets. 

\begin{table}[t]
\renewcommand\arraystretch{1.15}
    \caption{\textbf{Real-world Recordings.} Recording (Rec) duration is presented in the format of minimum/average/maximum duration. Ovl stands for overlap ratio.}
    % \footnotesize
    \label{tab:realdata}
    \centering
            \resizebox{\columnwidth}{!}{
            \begin{tabular}{cccccccc}
                \toprule
                Dataset & Split & \#Spk & \#Rec & Ovl (\%) & Duration (mins) \\
                \midrule
            \multirow{2}{*}{CALLHOME \cite{martin2001nist}} & Part 1 & 2$-$7 & 249 & 17.0 & 0.86/2.10/10.11   \\
                                ~ & Part 2 & 2$-$6 & 250 & 16.7 & 0.77/2.05/10.01  \\            
            \midrule
            \multirow{2}{*}{DIHARD II \cite{ryant2019seconddiharddiarizationchallenge}} & Dev & 1$-$10 & 192 & 9.8 & 0.45/7.44/11.62   \\
                                ~ & Test & 1$-$9 & 194 & 8.9 & 0.63/6.96/13.50  \\  
            \midrule
            \multirow{2}{*}{DIHARD III \cite{ryant21dh3_interspeech}} & Dev & 1$-$10 & 254 & 10.7 &  0.45/8.07/11.62  \\
                                ~ & Test & 1$-$9 & 259 & 9.4 & 0.63/7.65/13.50  \\
                                \midrule
            \multirow{3}{*}{AMI headset mix \cite{ami}} & Train & 3$-$5 & 136 & 13.4 & 7.97/35.59/90.25   \\
                                ~ & Dev & 4 & 18 & 14.1 & 15.73/32.22/49.50  \\
                                ~ & Test & 3$-$4 & 16 & 14.6 & 13.98/33.98/49.53 \\
                \bottomrule
            \end{tabular}
    }
    % \vspace{-1em}
\end{table}

CALLHOME consists of telephone conversations sampled at 8 kHz. DIHARD II and DIHARD III include recordings sampled at 16 kHz from diverse domains. AMI comprises meeting recordings sampled at 16 kHz. In our experiments, the recordings in DIHARD II, DIHARD III and AMI are down-sampled to 8 kHz to align with the sampling rate of the simulated dataset. 

\subsection{Model Configuration}
\label{sec_model_conf}
Following the feature extraction method presented in \cite{horiguchi2020end, horiguchi2022online}, the input audio features are set to 23-dimensional LogMels for every 10 ms, which are then concatenated with left and right 7 frames, and downsampled by a factor of 10, resulting in a 345-dimensional feature for every 0.1 s. 
% The configurations of the proposed model are shown in Table \ref{tab:mod_conf}. 
The number of stacked blocks in encoder and decoder are set to $N$=4 and $M$=2, respectively. For both the encoder and decoder, the number of attention heads and hidden units are set to 4 and 256, respectively. The feed-forward dimension is set to 1024 in the encoder and 2048 in the decoder. The convolution kernel size in the Conformer module is set to 16, with a zero-padding of 15 to ensure causality. In the  look-ahead module, the kernel size and padding are set to 19 and 9, respectively. Consequently, each frame attends to its previous and future 9 frames. Overall, the processing latency is $(1+9)\times 0.1 + 7 \times 0.01$ = 1.07 s. In the following, we approximate the overall delay as 1 s for presentation simplicity. As already mentioned, the decay factor $\gamma$ in Retention is set to 1. 

% \begin{table*}[t]
% \renewcommand\arraystretch{1.15}
%     \caption{\textbf{Model Configuration.} Kern\_1 / Pad\_1 and Kern\_2 / Pad\_2 stand for kernel size / padding in the Conformer module and Look-ahead module, respectively. FF\_enc and FF\_dec stand for feed-forward in the encdoer and decoder, respectively.}
%     % \footnotesize
%     \label{tab:mod_conf}
%     \centering
%     \resizebox{2\columnwidth}{!}{
%             \begin{tabular}{cccccccccc}
%                 \toprule
%                 \#Encoder Block & \#Decoder Block & \#Head & \#Hidden Unit & Kern\_1 / Pad\_1 & Kern\_2 / Pad\_2 & FF\_enc Dimension & FF\_dec Dimension & $\gamma$ \\
%                 \midrule
%                 4 & 2 & 4 & 256 & 16 / 15 & 19 / 9 & 1024 & 2048 & 1 \\
%                 \bottomrule
%             \end{tabular}
%             }
%     % \vspace{-1em}
% \end{table*}

\subsection{Training Setup}
\label{sec4c_train_setup}
The training process is divided into two stages: short-recording pre-training and long-segment adaptation. Pre-training is progressively conducted with linearly increased number of speakers, using the very large simulated dataset, and the length of training segments is set to 100 s.  Specifically, the model is trained in multiple steps with the subsets of Sim2spk, Sim\{1-4\}spk, Sim\{1-6\}spk and Sim\{1-8\}spk for 100, 50, 50 and 25 epochs, respectively. Training segments are randomly extracted from the original recordings in an on-the-fly manner.  For each step, the Noam optimizer \cite{vaswani2017attention} with 100,000 warm-up steps is utilized.

% Initially, the model is pre-trained on the Sim2spk dataset for 100 epochs. This is followed by pre-training on simulated mixtures of Sim\{1-4\}spk for 50 epochs. To adapt to scenarios with a large number of speakers, we further pre-train the model using simulated mixtures of Sim\{1-6\}spk and Sim\{1-8\}spk for 50 epochs and 25 epochs, respectively. 
% The resulting pre-trained models are referred to as LS-EEND-4SPK, LS-EEND-6SPK, and LS-EEND-8SPK. 
The pre-trained model cannot directly generalize to longer audio streams and real-world data. At the adaption stage, the pre-trained models are fine-tuned and adapted to specific applications/datasets. 
The fine-tuning process is also progressively conducted, but now with doubly increased audio lengths, using respective adaption data of each dataset. For real-world datasets, adaption is completed  simultaneously for longer audio streams and real-world data.  Audio length is increased from 200 s until it approximates the average recording duration of each dataset. 
% To handle varying audio stream lengths, the segment length is increased exponentially from 200 s until it approximates the average duration. 
Specifically, audio length is progressively set to 200 s and 400 s for the simulated dataset, to 200 s for CALLHOME, to 200 s and 400 s for DIHARD II and DIHARD III, to 200 s, 400 s, 800 s, and 1600 s for AMI.
The adaption segments are also randomly extracted from the original recordings in an on-the-fly manner. 
When training with long sequence, the chunk-wise recurrent paradigm described in Section \ref{sec_3b_model_arch} is used for the Retention modules, with a recurrent chunk length of 50 s. The Adam optimizer \cite{kingma2015adam} with a learning rate of $1 \times 10^{-5}$ is used. In our experiments, we find that the model fine-tuned using the BCE loss (according to the speaker appearance order) does not converge well when adapted to real-world recordings. 
% We believe this may be because the speaker order information, determined through label permutation, is difficult to transfer across different domains. 
Therefore, we utilize the PIT loss \cite{yu2017pit} for real-world datasets.

As mentioned in Section \ref{sec_infer}, during the inference stage, the model outputs activity probabilities corresponding to the maximum number of attractors. To enable the network automatically handle a flexible number of speakers below this upper bound, the termination marker (as illustrated in Fig. \ref{fig:transf_label}) is adopted. During training, for a training sample with $s$ ($s\leq S$) actual speakers, the $(s{+}1)$-th attractor is designated as the termination marker (with all zero labels), and the diarization loss is computed on the 0-th to the $(s{+}1)$-th attractor. By training the network to be aware of an explicit termination marker, then the network can automatically count active speakers at inference. 
% The zero label of termination marker serves to indicate that all active speakers have been successfully decoded by the network up to the $(s{+}1)$-th index. 
% This mechanism allows each sample to explicitly indicate the number of active speakers. 
% The posterior probabilities $\hat{\boldsymbol{y}}_{t}=(\hat{y}_{0,t},\hat{y}_{1,t}\ldots,\hat{y}_{s,t},\hat{y}_{\text{ter},t})$ are then estimated, and the diarization loss is computed with respect to the ground-truth labels according to Eq. (\ref{eq:dia_loss}).

\subsection{Evaluation Setup}
Diarization error rate (DER) is adopted as the evaluation metric, which is computed after performing the optimal speaker matching between the prediction and ground-truth label. The decision of diarization results is made using a threshold of 0.5. For the simulated data and CALLHOME, a collar tolerance of 0.25 s is used. For  DIHARD II, DIHARD III and AMI, no collar tolerance is used. Note that our system automatically outputs SAD information, which however may not be perfect. To ensure a fair comparison with diarization methods \cite{zhang2019fully, Fini2020supervised, zhang22_odyssey, yue22b_interspeech} that utilize the oracle SAD, we apply the SAD post-processing method presented in \cite{Horiguchi2022taslp} to the diarization results of the proposed model.

\section{Experimental Results}
This section presents the results. The fully trained model, namely pre-training to 8 speakers and fine-tuning with long audio, is considered as our final proposal, referred to as \textbf{LS-EEND (prop.)}. We also provide comparisons with some other configurations of the proposed model, which will be referred to as LS-EEND (*), where * specifies the configuration. For all comparison methods, the results are directly quoted from their papers or their subsequent papers by the same authors. Except that our previously proposed FS-EEND model \cite{liang2024framewise} is re-trained using data presented in this work.

\subsection{Results on Simulated Data}
% \subsubsection{DERs Evaluation} 
Table \ref{tab:DERs2} shows the  diarization performance on the simulated dataset. We compare with the series of EEND methods \cite{horiguchi2020end,han2021bw,Horiguchi2021ASRU,xue2021online2,Horiguchi2022taslp,horiguchi2022online,chentaslp2024} on the same test set. For the proposed method, the final model (with long-segment adaption) and the models at three pre-training steps, i.e. Sim\{1-4\}spk, Sim\{1-6\}spk and Sim\{1-8\}spk, are all evaluated. Note that, all the comparison methods were trained in a similar way as the proposed LS-EEND (Sim\{1-4\}spk) model. It can be seen that the proposed LS-EEND (Sim\{1-4\}spk) model achieves comparable performance with the advanced EEND-GLA-Large model for 1$\sim$4 speakers, which demonstrates the effectiveness of the proposed attractor decoder for learning consistent speaker attractors over a long time period in a streaming way.

As a common phenomenon presented in \cite{Horiguchi2021ASRU,Horiguchi2022taslp,horiguchi2022online}, the model trained with Sim\{1-4\}spk has a large performance degradation for the unseen case of 5 speakers. 
By increasing the number of training speakers to 6 and further to 8, the proposed model can handle more and more speakers, and achieves much lower DERs than comparison methods for 5$\sim$6 speakers. Moreover, along with the increase of training speakers, the proposed model performs even noticeably better for smaller numbers of speakers. Specifically, the Sim\{1-6\}spk model performs better for 1$\sim$4 speakers compared to the Sim\{1-4\}spk model, while the Sim\{1-8\}spk model performs better for 1$\sim$6 speakers compared to the Sim\{1-6\}spk model. 
% This is different from the results presented in \cite{Horiguchi2022taslp} that, for EEND-EDA, the performance of 2$\sim$4 speakers will be degraded when increasing the training speakers from \{1-4\}spk to \{1-5\}spk. The reason for this difference is possibly that, compared to the LSTM-based encoder-decoder attractor network in EEND-EDA, 
This demonstrates that the proposed self-attention-based attractor decoder is able to progressively accommodate a higher number of speakers/attractors, and learning for more speakers is even helpful for improving the distinction of attractors for smaller numbers of speakers. This property, namely the number of training speakers can be effectively increased  without sacrificing the performance of smaller numbers of speakers, is important, as it provides a simple yet effective way for the proposed model to handle a high (although not unlimited) number of speakers. Finally, the pre-trained Sim\{1-8\}spk is further improved after the adaption using long segment, especially for 7$\sim$8 speakers that have large recording duration.  
% achieves good performance for almost all the 
% 1$\sim$8 speakers, thence it will be used in the following experiments.    

Compared to our previous work FS-EEND \cite{liang2024framewise}, the present work not only reduces the computational complexity (will be shown later), but also largely improves the diarization performance, mainly by adopting the Conformer architecture, the Retention mechanism and the new training strategy. For example, the DERs is reduced from 14.93\% to 8.34\% for 4 speakers.

% We compare LS-EEND with recently proposed offline and online diarization systems. 
% The X-vector clustering model \cite{horiguchi2020end} trained with Sim\{1-4\}spk also serves as an offline baseline. We compare EEND-based offline/online systems, where EEND-GLA-Small and EEND-GLA-Large are equiped with four- and six-stacked encoders, respectively. The results in Table \ref{tab:DERs2} demonstrate that our system trained on Sim\{1-4\}spk outperforms BW-EDA-EEND and STB-based online methods. Surprisingly, our LS-EEND-8SPK model generally performs on par with offline EEND-based methods \cite{Horiguchi2021ASRU}. Additionally, LS-EEND-6SPK generally performs better than LS-EEND-4SPK on Sim\{1-4\}spk datasets and LS-EEND-8SPK generally performs better than LS-EEND-6SPK on Sim\{1-6\}spk datasets. This suggests that increasing the maximum number of speakers during the multi-step pre-training stage helps improve performance on datasets with the same number of speakers.

\begin{table}[t]
\renewcommand\arraystretch{1.15}
    \caption{DERs (\%) on the simulated dataset.}
    \footnotesize
    \label{tab:DERs2}
    \centering
    \tabcolsep0.04in 
\resizebox{\columnwidth}{!}{
\begin{tabular}{lccccccccc}
    \toprule
    \centering
    \multirow{2}*{Methods} & \multirow{1}{*}{latency } &  \multicolumn{8}{c}{Number of speakers}\\
    \cmidrule{3-10} 
    ~ & (s) & 1 & 2 & 3 & 4 & 5 & 6 & 7 & 8 \\
    \midrule
    \textbf{Offline} \\
    % \hspace{0.5em} X-vector clustering \cite{horiguchi2020end}               & - & 37.42 & 7.74 & 11.46 & 22.45 & 31.00 & 38.62 & - & - \\
    \hspace{0.5em} EEND-EDA \cite{horiguchi2020end, Horiguchi2022taslp, Horiguchi2021ASRU, horiguchi2022online}  & - & 0.15 & 3.19 & 6.60  & 9.26 & 23.11 & 34.97 & - & - \\
    \hspace{0.5em} EEND-GLA-Small \cite{Horiguchi2021ASRU}   & - & 0.25 & 3.53 & 6.79 & 8.98 & 12.44 & 17.98 & - & - \\
    \hspace{0.5em} EEND-GLA-Large \cite{Horiguchi2021ASRU}   & - & 0.09 & 3.54 & 5.74 & 6.79 & 12.51 & 20.42 \\
    \hspace{0.5em} AED-EEND-EE \cite{chentaslp2024}          & - & 0.07 & 2.45 & 4.71 & 7.04 & - & - & - & -\\
    \midrule
    \textbf{Online} \\
    \hspace{0.5em} BW-EDA-EEND \cite{han2021bw}            & 10 & 1.03 & 6.10 & 12.58 & 19.17 & - & - & - & - \\
    \hspace{0.5em} FS-EEND \cite{liang2024framewise}  & 1 & 0.44 & 4.94 & 11.01 & 14.93 & - & - & - & -\\   
    \hspace{0.5em} EEND-EDA + FW-STB \cite{xue2021online1, xue2021online2, horiguchi2022online} & 1 & 1.50 & 5.91 & 9.79 & 11.92 & 26.57 & 37.31 & - & - \\ 
    \hspace{0.5em} EEND-GLA-Small + BW-STB \cite{horiguchi2022online}           & 1 & 1.19 & 5.18 & 9.41 & 13.19 & 16.95 & 22.55 & - & - \\
    \hspace{0.5em} EEND-GLA-Large + BW-STB \cite{horiguchi2022online}           & 1 & 1.12 & 4.61 & 8.14 & 11.38 & 17.27 & 25.77 &- &-  \\ \vspace{-0.3cm} \\
    % \hspace{0.5em} LS-EEND (Sim\{1-4\}spk)           & 1  & 0.79 & 3.76 & 8.10 & 11.45 & \cellcolor{gray!30} 24.91 & - & - & - \\   
    \hspace{0.5em} LS-EEND (Sim\{1-4\}spk)           & 1  & 0.45 & 4.14 & 7.99 & 11.34 & \cellcolor{gray!30} 25.44 & - & - & - \\
    %32.93 & 42.33 & - & - \\
    % \hspace{0.5em} LS-EEND (Sim\{1-6\}spk)            & 1  & 1.03 & 3.26 & 6.95 & 9.83 & 13.92 & 18.55 & \cellcolor{gray!30} 27.04 & - \\
    \hspace{0.5em} LS-EEND (Sim\{1-6\}spk)            & 1  & 0.38 & 3.17 & 6.48 & 8.98 & 12.29 & 16.68 & \cellcolor{gray!30} 24.94 & - \\
    % \hspace{0.5em} LS-EEND  (Sim\{1-8\}spk)            & 1  & 1.69 & 3.34 & 6.37 & 8.88 & 12.40 & 17.14 & 22.84 & 29.10\\
    \hspace{0.5em} LS-EEND  (Sim\{1-8\}spk)            & 1  & \textbf{0.34} & 2.96 & 6.30 & 8.38 & 11.30 & 15.63 & 20.75 & 26.17\\
    \hspace{0.5em} LS-EEND  (prop.)             & 1  & \textbf{0.34} & \textbf{2.84} & \textbf{6.25} & \textbf{8.34} & \textbf{11.26} & \textbf{15.36} & \textbf{19.53} & \textbf{23.35}\\
    \bottomrule
\end{tabular}
}
% \vspace{-1em}
% \begin{tablenotes}
%     \item[1] $\hspace{1em} \dagger:$ progressive fine-tuning using 200s-$>$400s audio length. 
%     %\item[2] $\hspace{1em} *:$ new multi-step finetune results (200s-$>$400s) 
% \end{tablenotes}
\end{table}

\subsection{Results on Real-world Recordings}
In this section, we present the results on real-world datasets. The results of two SOTA offline end-to-end diarization methods, i.e. EEND-GLA-Large \cite{Horiguchi2021ASRU} and AED-EEND-EE \cite{chentaslp2024}, are also provided for reference. Scores are directly quoted from their papers when available. For AED-EEND-EE \cite{chentaslp2024}, among different variants, the best-performing system for each dataset is adopted. For comparison, the results of online diarization methods in the literature are also directly quoted from their papers. 

\subsubsection{Results on CALLHOME} 
Table \ref{tab:DERs3} shows the results on CALLHOME. 
% \addnote[ls-eend-pts]{1}{, where LS-EEND (PTS) represents the model trained using the proposed progressive training strategy (PTS). That is, fine-tuning with doubly increased audio length based on the short-segment pre-training LS-EEND (Sim\{1-8\}spk) model.}
The proposed model achieves new SOTA performance compared to other online methods. The performance advantages get more clear along with the increase of speakers. 
Considering that other online methods all used Sim\{1-4\}spk at the pre-training stage, we also provide the results (in grey) of the proposed model adapted from the pre-trained Sim\{1-4\}spk model for reference, which also show certain performance advantages. 
% it is also competitive with offline diarization methods. As the number of speakers in the pre-training process increases, the system's performance gradually improves, which is consistent with the results on simulated data. 

\begin{table}[t]
\renewcommand\arraystretch{1.15}
    \caption{DERs (\%) on CALLHOME.}
    \footnotesize
    \tabcolsep0.04in
    \label{tab:DERs3}
    \centering
\resizebox{\columnwidth}{!}{
\begin{tabular}{lccccccc}
    \toprule
    \multirow{2}{*}{Methods} & \multirow{1}{*}{latency} & \multicolumn{6}{c}{Number of speakers} \\
    \cmidrule{3-8} 
    ~ & (s) & 2 & 3 & 4 &5 & 6 & all \\
    \midrule
    \textbf{Offline} \\
    % \hspace{0.5em} SC-EEND \cite{Fujita2020NeuralSD}  & - & 9.57 & 14.00 & 21.14 & 31.07 & 37.06 & 15.75 \\
    % \hspace{0.5em} MTEAD \cite{cheng2023multi}        & - & - & - & - & - & - & 14.31 \\
    % \hspace{0.5em} EEDN-EDA \cite{horiguchi2020end,Horiguchi2022taslp} & - & 7.83 & 12.29 & 17.59 & 27.66 & 37.17 & 13.65 \\
    % \hspace{0.5em} EEND-VC-iGMM \cite{keisuke2022tight} & - & 8.6 & 12.6 & 16.1 & 27.5 & 26.9 & 13.3 \\
    % \hspace{0.5em} VBx \cite{landini2022bayesian} $\dagger$   & - & 9.44 & 13.89 & 16.05 & 13.87 & 24.73 & 13.28 \\
    % % \hspace{0.5em} AED-EEND-EE \cite{chentaslp2024}          & - & 6.93 & 11.92 & 17.12 & 28.22 & 31.97 & 12.91 \\
    % \hspace{0.5em} EEND-vector clust. \cite{kinoshita21_interspeech} & - & 7.94 & 11.93 & 16.38 & 21.21 & 23.10 & 12.49 \\    
    % \hspace{0.5em} EEND-GLA-Small \cite{Horiguchi2021ASRU} & - & 6.94 & 11.42 & 14.49 & 29.76 & 24.09 & 11.92 \\
    \hspace{0.5em} EEND-GLA-Large \cite{Horiguchi2021ASRU} & - & 7.11 & 11.88 & 14.37 & 25.95 & 21.95 & 11.84 \\
    \hspace{0.5em} AED-EEND-EE \cite{chentaslp2024}        & - & 5.69 & 9.81 & 12.44 & 23.35 & 21.72 & 10.08 \\
    \midrule
    \textbf{Online} \\
    \hspace{0.5em} BW-EEND-EDA \cite{han2021bw}              & 10 & 11.82 & 18.30 & 25.93 & - & - & - \\
    \hspace{0.5em} FS-EEND \cite{liang2024framewise}         & 1 & 10.1 & 14.6 & 21.2 & - & - & - \\
    \hspace{0.5em} EEND-EDA + FW-STB \cite{xue2021online1, xue2021online2, horiguchi2022online}   & 1 & 9.08 & 13.33 & 19.36 & 30.09 & 37.21 & 14.93 \\
    \hspace{0.5em} EEND-GLA-Small + BW-STB \cite{horiguchi2022online} & 1 & 9.01 & 12.73 & 19.45 & 32.26 & 36.78 & 14.80 \\
    \hspace{0.5em} EEND-GLA-Large + BW-STB \cite{horiguchi2022online} & 1 & 9.20 & 12.42 & 18.21 & 29.54 & 35.03 & 14.29 \\ \vspace{-0.3cm} \\
%     \hspace{0.5em} LS-EEND-4SPK (prop.)                 & 1  & 9.87 & 13.67 & 19.77 & 32.56 & 41.02 & 15.82 \\
    % \hspace{0.5em} LS-EEND-6SPK (prop.)                 & 1 & 8.85 & 
%12.35 & 18.28 & 34.32 & 33.96 & 14.33 \\
 %    \hspace{0.5em} LS-EEND-8SPK (prop.)                 & 1 & \textbf{8.13} & 
%\textbf{11.79} & 19.09 & 32.01 & \textbf{33.43} & 13.92 \\
%     \hspace{0.5em} LS-EEND-4SPK (prop.) *               & 1 & 9.98 & 14.48 & 18.33 & 33.09 & 49.39 & 16.25\\
 %    \hspace{0.5em} LS-EEND-6SPK (prop.) *               & 1 & 8.85 & 12.26 & 19.68 & 34.81 & 34.58 & 14.77 \\
    % \hspace{0.5em} LS-EEND-8SPK (prop.) *               & 1 & 8.62 & 12.12 & \textbf{16.78} & 32.33 & 34.20 & \textbf{13.78} \\

     % \color{gray} \hspace{0.5em} LS-EEND (Sim\{1-4\}spk)                 & \color{gray} 1 & \color{gray} 8.90 & \color{gray} 13.24 & \color{gray} 17.94 & \color{gray} 27.60 & \color{gray} 32.41 & \color{gray} 14.22 \\
     
     \color{gray} \hspace{0.5em} LS-EEND (Sim\{1-4\}spk)                 & \color{gray} 1 & \color{gray} 7.12 & \color{gray} 12.60 & \color{gray} 17.63 & \color{gray} 29.05 & \color{gray} 36.76 & \color{gray} 13.58 \\
     % \hspace{0.5em} LS-EEND-6SPK (prop.)                 & 1 & 7.74 & 12.01 & 16.65 & \textbf{25.13} & 31.89 & 12.98 \\
      % \hspace{0.5em} LS-EEND (prop.)                 & 1 & \textbf{6.64} & \textbf{11.34} & \textbf{16.57} & \textbf{26.38} & \textbf{30.28} & \textbf{12.33} \\
     \hspace{0.5em} {LS-EEND (prop.)}                 & 1 & \textbf{7.03} & \textbf{11.59} & \textbf{15.30} & \textbf{24.63} & \textbf{27.89} & \textbf{12.11} \\
    \bottomrule                                                                                                  
\end{tabular}
}
% \begin{tablenotes}
%     \item[1] $\hspace{1em} \dagger:$ Oracle SAD is used. 
%     %\item[2] $\hspace{1em} *:$ new multi-step finetune results (200s-$>$400s) 
% \end{tablenotes}
   % \vspace{-1em}
\end{table}

\subsubsection{Results on DIHARD II and DIHARD III}
Table \ref{tab:DERs5} and \ref{tab:DERs6} show the results on DIHARD II and DIHARD III, respectively. These two datasets are challenging in the sense that up to 9 speakers can be involved in the test recordings. The proposed model achieves the best performance among online methods under most conditions, which again verifies that the proposed model can simply handle a large number speakers by progressively increasing the training speakers. 
For the case with oracle SAD on DIHARD III, the proposed model performs better than EEND-GLA-Small and Conformer OTS-VAD, but comparable or worse than their variants, i.e. EEND-GLA-Large and ResNet OTS-VAD. The encoder architecture of the proposed model, i.e. four stacks of Conformer blocks, is more similar to the one of EEND-GLA-Small and Conformer OTS-VAD. These results show that leveraging a larger encoder as in EEND-GLA-Large or a CNN encoder as in ResNet OTS-VAD is possibly more suitable for some scenarios.

% which  The DIHARD II results are shown  in . LS-EEND (4SPK) outperforms other online methods when the number of speakers is low, but not when the number of speakers is high. This is because the attractors corresponding to the high-index speakers are not trained during the pre-training phase and are not sufficiently trained on the adaptation set alone. This issue can be addressed by increasing the number of speakers in the pre-training set, as evidenced by the significant improvement in DER from 47.32\% to 45.26\% with LS-EEND. When no oracle information is used, our system outperforms both end-to-end STB-based methods and the cascaded overlap-aware speaker embedding \cite{coria2021overlap}. With oracle SAD, our system achieves the best overall performance compared to other methods.

% The DIHARD III results shown in Table \ref{tab:DERs6} are similar to those of DIHARD II. When using oracle SAD, our system outperforms the conventional cascaded methods and is only slightly worse than the best method EEND-GLA-Large + BW-STB \cite{horiguchi2022online} by 0.51\%. Given that EEND-GLA-Large + BW-STB employs six encoder blocks, we anticipate that the performance of LS-EEND will improve further with additional embedding encoder blocks. Moreover, it outperforms other online methods when no oracle information is used, which is the primary benchmark for end-to-end methods.

\begin{table}[t]
\renewcommand\arraystretch{1.15}
    \caption{DERs (\%) on DIHARD II.}
    \footnotesize
    \label{tab:DERs5}
    \centering
\resizebox{\columnwidth}{!}{
\begin{tabular}{lccccc}
    \toprule
    \multirow{2}{*}{Methods} & \multirow{1}{*}{latency} & \multicolumn{3}{c}{Number of speakers} \\
    \cmidrule{3-5} 
    ~ & (s) & $\leq$ 4 & $\geq$ 5 & all \\
    \midrule
    \textbf{Offline} \\
    % \hspace{0.5em} EEND-EDA \cite{horiguchi2020end,Horiguchi2022taslp}  & - & 22.09 & 47.66 & 30.07 \\
    % \hspace{0.5em} EEND-GLA-Small \cite{Horiguchi2021ASRU}  & - & 22.24 & 44.92 & 29.31 \\
    \hspace{0.5em} EEND-GLA-Large \cite{Horiguchi2021ASRU}  & - & 21.40 & 43.62 & 28.33 \\
    % \hspace{0.5em} BUT system \cite{landini2020but}   & - & 21.34 & 39.85 & 27.11 \\
    % \hspace{0.5em} VBx + overlap-aware resegmentation \cite{bredin21_interspeech} & - & 21.41 & 36.93 & 26.25 \\
    \hspace{0.5em} AED-EEND-EE \cite{chentaslp2024}         & - & - & - & 24.64 \\
    \midrule
    \textbf{Online} \\                                                  
    \hspace{0.5em} Overlap-aware speaker embeddings \cite{coria2021overlap} & 1 & 27.00 & 52.62 & 34.99 \\
    \hspace{0.5em} EEND-EDA + FW-STB \cite{xue2021online1, xue2021online2, horiguchi2022online} & 1 & 25.63 & 50.45 & 33.37 \\
    \hspace{0.5em} EEND-GLA-Small + BW-STB \cite{horiguchi2022online} & 1 & 23.96 & 48.06 & 31.47 \\
    \hspace{0.5em} EEND-GLA-Large + BW-STB \cite{horiguchi2022online} & 1 & 22.62 & 47.06 & 30.24 \\ \vspace{-0.3cm} \\
    % \hspace{0.5em} LS-EEND-4SPK (prop.)                & 1 & 24.37 & 52.64 & 34.00 \\
    % \hspace{0.5em} LS-EEND-6spk (prop.)                & 1 & 22.80 & 48.58 & 30.95 \\
    % \hspace{0.5em} LS-EEND-8SPK (prop.)                & 1 & 23.41 & 47.65 & 31.23 \\
    % \hspace{0.5em} LS-EEND-4SPK (prop.) *              & 1 & 23.26 & 51.29 & 32.25 \\
    % \hspace{0.5em} LS-EEND-6spk (prop.) *              & 1 & \textbf{22.68} & 48.56 & 30.75 \\
    % \hspace{0.5em} LS-EEND-8SPK (prop.) *              & 1 & 23.19 & \textbf{46.48} & \textbf{30.45} \\
    \color{gray} \hspace{0.5em} LS-EEND (Sim\{1-4\}spk)           & \color{gray} 1 & \color{gray} 21.56 & \color{gray} 47.32 & \color{gray} 29.59 \\
    % \hspace{0.5em} LS-EEND-6spk (prop.)                & 1 & 21.35 & 45.91 & 29.00 \\
    % \hspace{0.5em} LS-EEND (prop.)                & 1 & \textbf{21.01} & \textbf{45.26} & \textbf{28.57} \\
    \color{gray} \hspace{0.5em} LS-EEND (long pre-train)                & \color{gray} 1 & \color{gray} 20.74 & \color{gray} 43.88 & \color{gray} 27.95 \\
    \hspace{0.5em} {LS-EEND (prop.)}               & 1 & \textbf{20.52} & \textbf{43.17} & \textbf{27.58} \\

    \textbf{Online (with oracle SAD)} \\
    \hspace{0.5em} UIS-RNN \cite{zhang2019fully}           & 1 & - & - & 30.9 \\
    \hspace{0.5em} UIS-RNN-SML \cite{Fini2020supervised}   & 1 & - & - & 27.3 \\
    \hspace{0.5em} Core sample selection \cite{yue22b_interspeech} & 1 & - & - & 23.1 \\
    \hspace{0.5em} EEND-EDA + FW-STB \cite{xue2021online1, xue2021online2, horiguchi2022online} & 1 & 16.56 & 42.58 & 24.67 \\
    \hspace{0.5em} EEND-GLA-Small + BW-STB \cite{horiguchi2022online} & 1 & 15.29 & 40.85 & 23.26 \\
    \hspace{0.5em} EEND-GLA-Large + BW-STB \cite{horiguchi2022online} & 1 & 13.55 & 40.39 & 21.92 \\ \vspace{-0.3cm} \\
    % \hspace{0.5em} LS-EEND-4SPK (prop.)                & 1 & 16.11 & 44.17 & 26.10 \\
    % \hspace{0.5em} LS-EEND-6SPK (prop.)                & 1 & 15.46 & 40.33 & 23.30 \\
    % \hspace{0.5em} LS-EEND-8SPK (prop.)                & 1 & 15.57 & 39.59 & 23.39 \\
    % \hspace{0.5em} LS-EEND-4SPK (prop.) *              & 1 & 16.19 & 42.83 & 24.49 \\
    % \hspace{0.5em} LS-EEND-6SPK (prop.) *              & 1 & \textbf{15.14} & 40.75 & 23.12 \\
    % \hspace{0.5em} LS-EEND-8SPK (prop.) *              & 1 & 15.32 & \textbf{39.24} & \textbf{22.78} \\
    \color{gray} \hspace{0.5em} LS-EEND (Sim\{1-4\}spk)     & \color{gray} 1 & \color{gray} 14.82 & \color{gray} 40.91  & \color{gray} 22.95 \\
    % \hspace{0.5em} LS-EEND-6SPK (prop.)                & 1 & 14.23 & 39.60 & 22.13 \\
    % \hspace{0.5em} LS-EEND (prop.)                & 1 & 13.85 & \textbf{38.59} & \textbf{21.57} \\
   \color{gray} \hspace{0.5em} LS-EEND (long pre-train)     & \color{gray} 1 & \color{gray} {13.00} & \color{gray} 37.48 & \color{gray} 20.63 \\
    \hspace{0.5em} {LS-EEND (prop.)}              & 1 & \textbf{13.05} & \textbf{37.18} & \textbf{20.57} \\    
    \bottomrule                                                                                                  
\end{tabular}
}
% \begin{tablenotes}
%     % \item[1] $\hspace{1em} * :$ 200s-$>$400s fine-tune using 100s simu model
%     \item[1] \parbox[t]{\linewidth}{\hangindent=1em \hangafter=1 $ \hspace{1em} \dagger:$ fine-tuning using 400s audio length on the 100s-$>$200s-$>$400s pre-trained model.}
% \end{tablenotes}
   % \vspace{-1em}
\end{table}

\begin{table}[t]
\renewcommand\arraystretch{1.15}
    \caption{DERs (\%) on DIHARD III.}
    \footnotesize
    \label{tab:DERs6}
    \centering
\resizebox{\columnwidth}{!}{
\begin{tabular}{lccccc}
    \toprule
    \multirow{2}{*}{Methods} & \multirow{1}{*}{latency} & \multicolumn{3}{c}{Number of speakers} \\
    \cmidrule{3-5} 
    ~ & (s) & $\leq$ 4 & $\geq$ 5 & all \\
    \midrule
    \textbf{Offline} \\
    % \hspace{0.5em} VBx + overlap handling \cite{horiguchi2021thehitachi}  & - & 16.38 & 42.51 & 21.47 \\
    % \hspace{0.5em} VBx + overlap-aware resegmentation \cite{bredin21_interspeech} & - & 15.32 & 35.87 & 19.33 \\
    % \hspace{0.5em} EEND-EDA \cite{horiguchi2020end, Horiguchi2022taslp}  & - & 15.55 & 48.30  & 21.94  \\
    % \hspace{0.5em} EEND-GLA-Small \cite{Horiguchi2021ASRU}               & - & 14.39 & 44.32 & 20.23 \\
    \hspace{0.5em} EEND-GLA-Large \cite{Horiguchi2021ASRU}               & - & 13.64 & 43.67 & 19.49 \\
    \midrule
    \textbf{Online} \\                                                  
    \hspace{0.5em} Overlap-aware speaker embeddings \cite{coria2021overlap}  & 1 & 21.07 & 54.28 & 27.55\\
    \hspace{0.5em} EEND-EDA + FW-STB \cite{xue2021online1, horiguchi2022online, xue2021online2} & 1 & 19.00 & 50.21 & 25.09 \\
    \hspace{0.5em} EEND-GLA-Small + BW-STB \cite{horiguchi2022online} & 1 & 15.87 & 47.27 & 22.00 \\
    \hspace{0.5em} EEND-GLA-Large + BW-STB \cite{horiguchi2022online} & 1 & 14.81 & 45.17 & 20.73 \\ \vspace{-0.3cm} \\
 %    \hspace{0.5em} LS-EEND-4SPK (prop.)                & 1 & 16.67  & 52.86 & 24.07
 % \\
 %    \hspace{0.5em} LS-EEND-6spk (prop.)                & 1 & \textbf{15.02} & 50.89 & 22.29
 % \\ 
 %    \hspace{0.5em} LS-EEND-8SPK (prop.)                & 1 & 15.77 & \textbf{47.21}  &  21.96 \\
 %    \hspace{0.5em} LS-EEND-4SPK (prop.) *              & 1 & 16.18 & 51.22 & 23.00 \\
 %    \hspace{0.5em} LS-EEND-6SPK (prop.) *              & 1 & 15.63 & 49.54 & 22.23 \\
 %    \hspace{0.5em} LS-EEND-8SPK (prop.) *              & 1 & 15.16 & 48.71 & \textbf{21.69} \\
    \color{gray} \hspace{0.5em} LS-EEND (Sim\{1-4\}spk)     & \color{gray} 1 & \color{gray} 14.56 & \color{gray} 46.01 & \color{gray} 20.69 
  \\
 %    \hspace{0.5em} LS-EEND-6spk (prop.)                & 1 & 15.15 & 45.01 & 20.97
 % \\ 
    % \hspace{0.5em} LS-EEND (prop.)                & 1 & \textbf{14.35} & \textbf{44.63} & \textbf{20.25} \\
  \color{gray}  \hspace{0.5em} LS-EEND (long pre-train)      & \color{gray} 1 & \color{gray} 14.35 & \color{gray} 42.57 & \color{gray} 19.85 \\
    \hspace{0.5em} {LS-EEND (prop.)}               & 1 & \textbf{13.96} & \textbf{42.98} & \textbf{19.61} \\
    
    \textbf{Online (with oracle SAD)} \\
    \hspace{0.5em} Zhang et al. \cite{zhang22_odyssey} & 0.5 & - & - & 19.57 \\
    \hspace{0.5em} Core sample selection \cite{yue22b_interspeech} & 1 & - & - &  19.3 \\
        \hspace{0.5em} EEND-EDA + FW-STB \cite{xue2021online1, xue2021online2, horiguchi2022online} & 1 & 12.80 & 42.46 & 18.58 \\
    \hspace{0.5em} EEND-GLA-Small + BW-STB \cite{horiguchi2022online} & 1 &  9.91 & 40.21 & 15.82\\
    \hspace{0.5em} EEND-GLA-Large + BW-STB \cite{horiguchi2022online} & 1 & \textbf{8.85} & 38.86 &14.70 \\
        \hspace{0.5em} Conformer OTS-VAD \cite{wang2023endtoendonlinespeakerdiarization} * & 0.8 & - & - & 16.33 \\
    \hspace{0.5em} ResNet OTS-VAD \cite{wang2023endtoendonlinespeakerdiarization} * & 0.4 & - & - &  \textbf{13.65} \\ \vspace{-0.3cm} \\
    % \hspace{0.5em} LS-EEND-4SPK (prop.)                & 1 & 11.43 & 44.18 & 18.48 \\
    % \hspace{0.5em} LS-EEND-6SPK (prop.)                & 1 & 10.52  & 43.63 & 17.45 \\
    % \hspace{0.5em} LS-EEND-8SPK (prop.)                & 1 & 10.74 & \textbf{38.86} & 16.24 \\
    % \hspace{0.5em} LS-EEND-4SPK (prop.) *              & 1 & 11.63 & 42.14 & 17.57 \\
    % \hspace{0.5em} LS-EEND-6SPK (prop.) *              & 1 & 11.49 & 41.82 & 17.40 \\
    % \hspace{0.5em} LS-EEND-8SPK (prop.) *              & 1 & 10.36 & 41.09 & 16.35 \\

    \color{gray} \hspace{0.5em} LS-EEND (Sim\{1-4\}spk)   & \color{gray} 1 & \color{gray} 10.35 & \color{gray} 39.29 & \color{gray} 15.99 \\
    % \hspace{0.5em} LS-EEND-6SPK (prop.)                & 1 & 10.88 & 38.95 & 16.35 \\
    % \hspace{0.5em} LS-EEND (prop.)                & 1 & 9.63 & \textbf{38.30} & 15.21 \\
    \color{gray} \hspace{0.5em} LS-EEND (long pre-train)      & \color{gray} 1 & \color{gray} 9.48 & \color{gray} {36.69} & \color{gray} 14.78 \\
    \hspace{0.5em} {LS-EEND (prop.)}               & 1 & 9.23 & \textbf{37.05} & 14.65 \\
    \bottomrule                                                                                                  
\end{tabular}
}
\begin{tablenotes}
    % \item[2] $\hspace{1em} *:$ new multi-step finetune results (200s-$>$400s-$>$800s-$>$1600s) 
    % \item[1] \parbox[t]{\linewidth}{\hangindent=1em \hangafter=1 $ \hspace{1em} \dagger:$ fine-tuning using 400s audio length on the 100s-$>$200s-$>$400s pre-trained model.}

     \item[2]  $ \hspace{1em} *:$ with accumulation-based inference process \cite{wang2023endtoendonlinespeakerdiarization}.
\end{tablenotes}
   % \vspace{0.7em}
\end{table}

In the proposed training strategy, model adaption with real-world datasets is always based on the short-segment pre-trained model. One alternative model for real-world adaption could be the one already fine-tuned with simulated long segments, e.g. the last model in Table~\ref{tab:DERs2}. The results of this setting (long pre-train) are also provided in Table \ref{tab:DERs5} and \ref{tab:DERs6}, where  adaption is performed directly using 400 s real-world data. It can be seen that this alternative performs slightly worse than the proposed one, which means it is better to conduct model adaption starting with short segments. 

\subsubsection{Results on AMI}
\label{sec5c1_ami}
\begin{figure}[t]
\centering
\includegraphics[width=\linewidth]{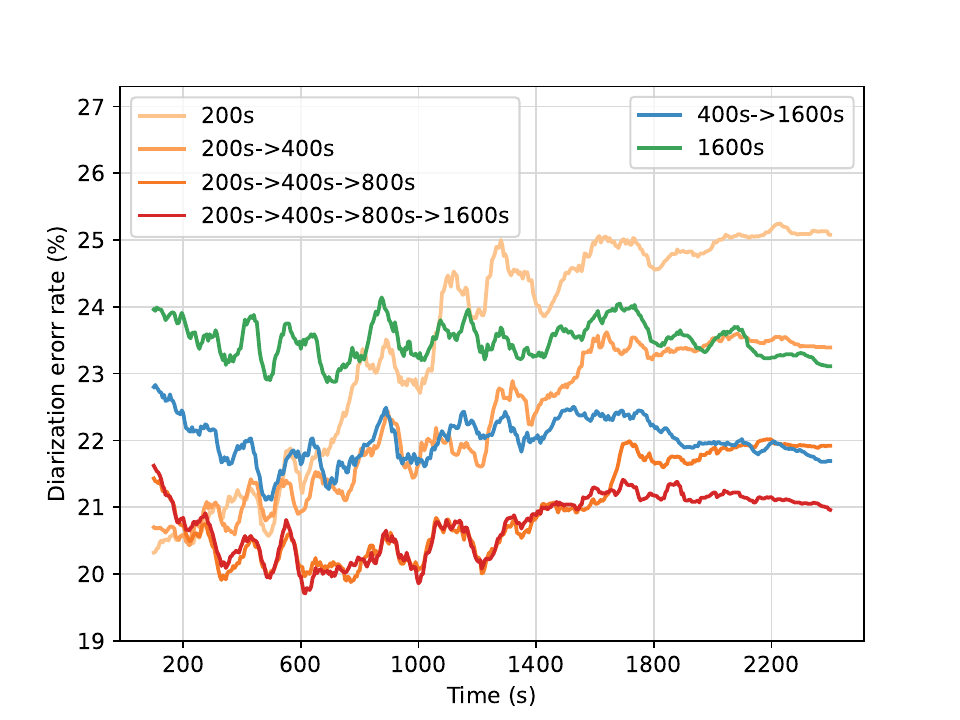}
\caption{DER curve with respect to audio length, for the AMI Dev set. `-$>$' denotes the next step of adaptation.}
\label{fig:der_dev}
\end{figure}

AMI has an average recording duration of 33 minutes, which pose significant challenges especially for online diarization systems to maintain the speaker consistency over such a long time period. 
As discussed in Section \ref{sec_train_strategy}, 
the block-wise online diarization methods \cite{horiguchi2022online, xue2021online2} can simply process long audio streams block by block, however the difficulty of resolving the speaker permutation ambiguity between chunks gets higher for them with the increase of audio length.
This work aims to leverage one single model to process long audio streams frame by frame, and speaker embeddings and attractors are kept consistent along the entire stream. This is achieved by training with long segments in a progressive way. To analyze the effectiveness of this training strategy, 
% the chunk-wise recurrent representation of Retention alleviates $O(n^2)$ complexity introduced by the parallel paradigm, thereby facilitating adaptation to long recordings. 
DER of the proposed model w.r.t audio length with different training strategies is plotted in Fig. \ref{fig:der_dev}. Note that the DER curves were smoothed using the Savizky-Golay filter \cite{savgol1964analytical} for visualization purpose. 
% We remind that these models are obtained by fine-tuning the Sim\{1-8\}spk pre-trained model using AMI data with different audio lengths. 
It can be seen that the 200s model suffers from the length extrapolation problem, namely the performance decreases when test audio length is larger than training audio length. Directly fine-tuning with 1600 s data solves the length extrapolation problem, but degrades the performance of short audio part. The proposed progressive training strategy, i.e. 200s-$>$400s-$>$800s-$>$1600s, gradually improve the performance of longer audio without degrading the performance of short audio part. This demonstrates that there does exist an optimization point being optimal for both the short and long audio parts, and the proposed training strategy is effective for reaching such optimal point. When the progress step is set to a larger value, the 400s-$>$1600s model has a prominent performance decrease. 
% segments converges to a  local optimum. Inspired by these observations, we explore multi-step adaptation by exponentially increasing the segment length from 200 s to 1600 s (approximating the average recording duration). It can be observed that multi-step adaptation maintains high performance on short speech and progressively reduce diarization error on long speech. We also investigate a two-step adaptation from 400 s to 1600 s (the blue curve in Fig. \ref{fig:der_dev}). The DER curve indicates that while it performs well on long speech, there is a slight performance degradation on short speech. Therefore, we propose an exponential increase with a base of 2 up to the average duration in this work.
Table \ref{tab:DERs4} shows the comparison results on AMI. We can see that the proposed model outperforms other online systems by a large margin. 

% demonstrate that our system outperforms other online diarization systems that do not use any oracle information, as well as some cascaded online methods using oracle SAD. Moreover, the gap between our system and offline EEND-EDA is less than two percentage points (23.64\% vs. 21.93\% on Dev and 21.51\% vs. 21.56\% on Eval).

\begin{table}[t]
\renewcommand\arraystretch{1.15}
    \caption{DERs (\%) on AMI. }
    \footnotesize
    \label{tab:DERs4}
    \centering
\resizebox{\columnwidth}{!}{
\begin{tabular}{lcccccc}
    \toprule
    \multirow{2}{*}{Methods} & \multirow{1}{*}{latency} & \multirow{2}{*}{Dev} & \multirow{2}{*}{Eval} \\
    ~ & (s) & ~ & ~  \\
    \midrule
    \textbf{Offline} \\
    % \hspace{0.5em} X-vec AHC + VB + OVL \cite{Horiguchi2021end, Garcia2020odyssey}             & - & - & 28.15 \\
    % \hspace{0.5em} SA-EEND \cite{fujita2019end2, Horiguchi2022taslp}    & - & 31.66 & 27.70 \\
    % \hspace{0.5em} Transcribe-to-Diarize \cite{kanda2022transcibe}  & - & 23.51 & 24.43 \\
    % \hspace{0.5em} Multi-Class Spec-Clustering \cite{raj2021multiclass} $\dagger$ & - & - & 23.60 \\
    % \hspace{0.5em} CmpEm + Overlap Detector \cite{li2021compositional}   & - & - & 22.93 \\
    % \hspace{0.5em} EENDA-EDA \cite{horiguchi2020end, Horiguchi2022taslp}  & - & 21.93 & 21.56 \\
    % \hspace{0.5em} NSD-MA-MSE \cite{he2023ansdmamse}    & - & 16.71 & 16.95 \\
    \hspace{0.5em} AED-EEND-EE + Conformer \cite{chentaslp2024}     & - & 13.63 & 13.00 \\
    \midrule
    \textbf{Online} \\
    \hspace{0.5em} Overlap-aware speaker embeddings \cite{coria2021overlap} & 1 & - & 30.4 \\
    \hspace{0.5em} SSep AMI + VAD E2E \cite{gruttadauria2024online} & 5 & - & 27.2 \\
    % \hspace{0.5em} LS-EEND-4SPK (prop.)          & 1 & 24.67 & 26.59 \\
    % \hspace{0.5em} LS-EEND-6SPK (prop.)          & 1 & \textbf{23.36} & 23.28 \\
    % \hspace{0.5em} LS-EEND-8SPK (prop.)          & 1 & 23.69 & 23.68 \\
    % \hspace{0.5em} LS-EEND-4SPK (prop.)        & 1 & 25.06 & 24.13 \\
    % \hspace{0.5em} LS-EEND-6SPK (prop.)        & 1 & 24.87 & 23.56 \\
    % \hspace{0.5em} LS-EEND (prop.)        & 1 & \textbf{23.64} & \textbf{21.51} \\
    \hspace{0.5em} {LS-EEND (prop.)}        & 1 & \textbf{20.97} & \textbf{20.76} \\
    \textbf{Online (with oracle SAD)} \\
    \hspace{0.5em} Conservative SD \cite{kwon2023absolute} & 0.5 & - & 22.88 \\
    \hspace{0.5em} Core sample selection \cite{yue22b_interspeech}             & 1 & - & 19.0 \\
    % \hspace{0.5em} LS-EEND-4SPK (prop.)          & 1 & 19.74 & 21.98 \\
    % \hspace{0.5em} LS-EEND-6SPK (prop.)          & 1 & 18.52 & 18.78 \\
    % \hspace{0.5em} LS-EEND-8SPK (prop.)          & 1 & 19.10 & 19.17 \\
    % \hspace{0.5em} LS-EEND-4SPK (prop.)        & 1 & 20.21 & 19.67 \\
    % \hspace{0.5em} LS-EEND-6SPK (prop.)        & 1 & 19.40 & 19.40 \\
    % \hspace{0.5em} LS-EEND (prop.)        & 1 &  \textbf{18.34} & \textbf{17.39} \\
    \hspace{0.5em} {LS-EEND (prop.)}        & 1 &  \textbf{16.74} & \textbf{16.66} \\
    \bottomrule                                                                                                  
\end{tabular}
}
% \begin{tablenotes}
%     \item[1] $\hspace{1em} \dagger:$ Oracle SAD is used. 
%     % \item[2] $\hspace{1em} *:$ new multi-step finetune results (200s-$>$400s-$>$800s-$>$1600s) 
% \end{tablenotes}
   % \vspace{1em}
\end{table}

\begin{figure}[t]
    \centering
    
    % 第一行
    \begin{subfigure}{0.495\linewidth}
        \centering
        \includegraphics[width=\linewidth]{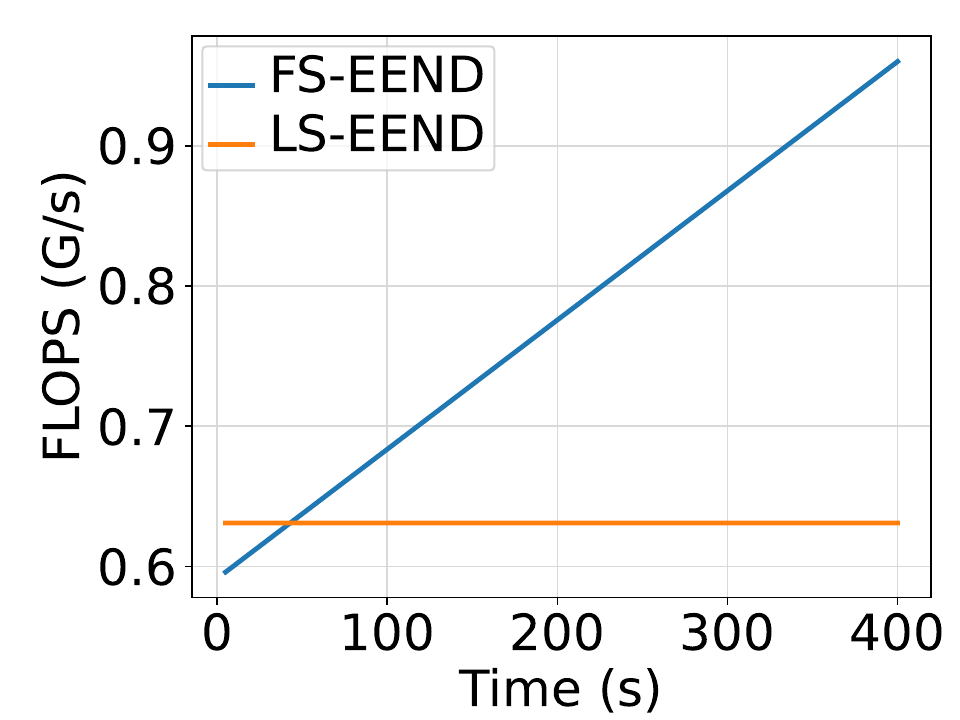}
        \captionsetup{labelformat=empty} % 取消标号
        \caption{(a) FLOPS (G/s)} % 使用英文替代中文标题
        \label{fig:subfig51}
    \end{subfigure}
    % \hfill
    \hspace*{-0.3cm}
    \begin{subfigure}{0.495\linewidth}
        \centering
        \includegraphics[width=\linewidth]{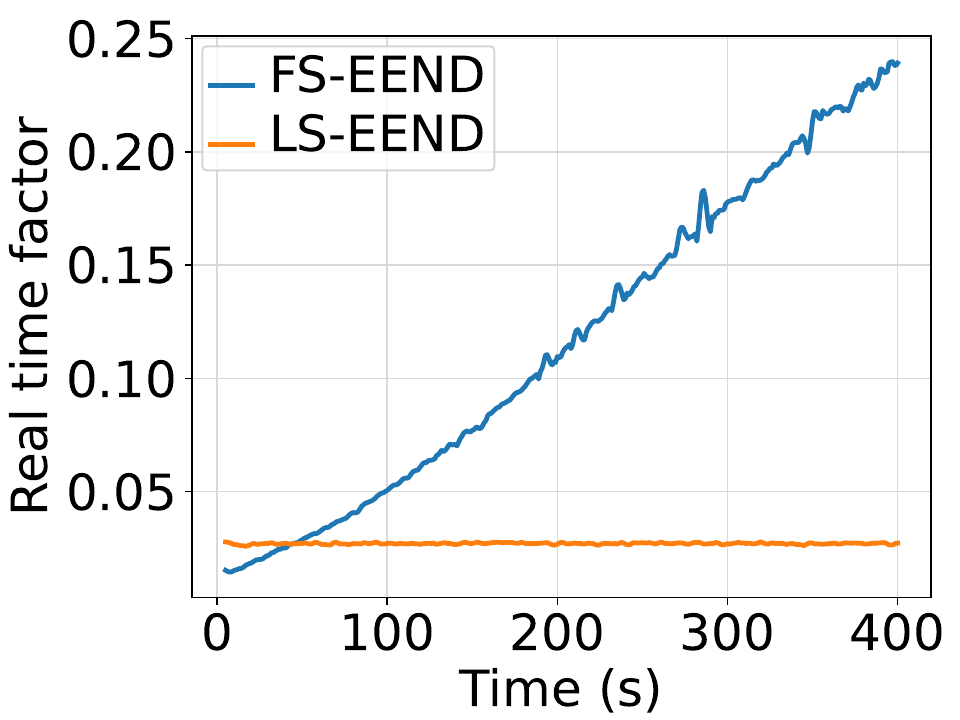}
        \captionsetup{labelformat=empty} % 取消标号
        \caption{(b) RTF } % 使用英文替代中文标题
        \label{fig:subfig52}
    \end{subfigure}

    % \vspace{0.08cm} % 行间距，可以根据需要调整

    \caption{FLOPS and RTF comparison between FS-EEND \cite{liang2024framewise} with self-attention and the proposed LS-EEND with Retention.} % 整个图的英文标题
    \label{fig:rtf}
\end{figure}

\setlength{\tabcolsep}{2pt}
\begin{table}[t]
\renewcommand\arraystretch{1.15}
    \caption{Real time factor comparison.}
    % \footnotesize
    \tabcolsep0.001in
    \label{tab:rtf}
    \centering
            \resizebox{\columnwidth}{!}{
            \begin{tabular}{cccccccc}
                \toprule
                Method & Device & \#Thread & RTF  \\
                \midrule
            Core sample selection \cite{yue22b_interspeech} $\dagger$ & NVIDIA Geforce RTX
3090 GPU & 1 & 0.10 \\
            % Overlap-aware speaker embeddings \cite{coria2021overlap} & CPU & 20 & 0.33 \\
            Zhang et al. \cite{zhang22_odyssey} $\dagger$ & Intel(R) Xeon(R) CPU E5-2630 v4 @ 2.20GHz & 1 & 0.10 \\
            EEND-GLA-Small + BW-STB \cite{horiguchi2022online} $\dagger$ & Intel Xeon Gold
6132 CPU @ 2.60 GHz & 7 & 0.16 \\
ResNet OTS-VAD \cite{wang2023endtoendonlinespeakerdiarization} $\dagger$ & NVIDIA Geforce RTX 3090 GPU & - & 0.16 \\
            EEND-EDA + FW-STB \cite{xue2021online2}*    &  AMD EPYC 7742 64-Core CPU @
2.25 GHz   & 1 & 0.247 \\
            LS-EEND (prop.) &  AMD EPYC 7742 64-Core CPU @
2.25 GHz & 1 & \textbf{0.028} \\
                \bottomrule
            \end{tabular}
    }
\begin{tablenotes}
    \item[1] $\hspace{1em} \dagger:$ RTF values are quoted from their original papers. 
    \item[2] $\hspace{1em} *:$ run inference on the same platform as the proposed model. 
\end{tablenotes}
    % \vspace{-1em}
\end{table}

\subsection{Computational Efficiency Analysis}
To assess the computational efficiency, we measure the floating point operations per second (FLOPS) \footnote{FLOPS (G/s) is calculated by counting the number of floating point operations and then dividing by the sequence length. We use the official tool provided by PyTorch (torch.utils.flop\_counter.FlopCounterMode) for FLOPS computation.} and real time factor (RTF). Our finally proposed model, i.e. the model trained with Sim\{1-8\}spk, is evaluated. The calculation is performed on an AMD EPYC 7742 64-Core CPU @ 2.25 GHz using a single thread, without using any GPU.

We first compare the complexity between the proposed model and our previous model FS-EEND \cite{liang2024framewise}, which use Retention and causal self-attention, respectively. At inference, Retention uses the recurrent paradigm. The FLOPS and RTF as a function of audio length are shown in Fig. \ref{fig:rtf}. 
% The inference process of LS-EEND follows its recurrent paradigm, and in FS-EEND \cite{liang2024framewise}, the attention for each frame is calculated with all preceding frames. As illustrated in Fig. \ref{fig:rtf},
Due to the quadratic temporal complexity of self-attention, FS-EEND roughly has a linear increase in FLOPS/RTF with audio length. In contrast, due to the linear complexity of Retention, the FLOPS/RTF of the proposed model remains constant regardless of the audio length, which makes the model suitable for processing very long audio streams.

Additionally, we compare RTFs with several SOTA online methods in Table \ref{tab:rtf}. The cascaded clustering-based methods \cite{yue22b_interspeech, zhang22_odyssey} usually exhibit a high RTF. \cite{zhang22_odyssey} improves the computational efficiency by utilizing restored checkpoint state in agglomerative hierarchy clustering (AHC). In block-wise online diarization methods \cite{horiguchi2022online,wang2023endtoendonlinespeakerdiarization}, normally a large block length is required to ensure the processing accuracy, while a small block shift is required to have a small processing latency. As a result, due to the large block overlap, the computational efficiency becomes lower, as every frame needs to be processed $\frac{\text{block length}}{\text{block shift}}$ times.   
% To resolve speaker permutation across blocks, the end-to-end method \cite{horiguchi2022online} needs to compute speaker embedding $\frac{\text{buffer size}}{\text{chuck size}}$ times per frame due to the overlap of blocks. 
In contrast, the proposed model processes once per frame, resulting in the RTF about an order of magnitude lower than that of EEND-EDA + FW-STB when using the same computing device, and several times lower than other methods although some of them use GPUs or multiple threads.

% \begin{figure}[t]
% \centering
% \includegraphics[width=0.9\linewidth]{figs/flops.pdf}
% \caption{FLOPS comparison on Sim5spk dataset.}
% \label{fig:rtf}
% \end{figure}

% \section{Algorithms}

% \begin{algorithm}[H]
% \caption{Weighted Tanimoto ELM.}\label{alg:alg1}
% \begin{algorithmic}
% \STATE 
% \STATE {\textsc{TRAIN}}$(\mathbf{X} \mathbf{T})$
% \STATE \hspace{0.5cm}$ \textbf{select randomly } W \subset \mathbf{X}  $
% \STATE \hspace{0.5cm}$ N_\mathbf{t} \gets | \{ i : \mathbf{t}_i = \mathbf{t} \} | $ \textbf{ for } $ \mathbf{t}= -1,+1 $
% \STATE \hspace{0.5cm}$ B_i \gets \sqrt{ \textsc{max}(N_{-1},N_{+1}) / N_{\mathbf{t}_i} } $ \textbf{ for } $ i = 1,...,N $
% \STATE \hspace{0.5cm}$ \hat{\mathbf{H}} \gets  B \cdot (\mathbf{X}^T\textbf{W})/( \mathbb{1}\mathbf{X} + \mathbb{1}\textbf{W} - \mathbf{X}^T\textbf{W} ) $
% \STATE \hspace{0.5cm}$ \beta \gets \left ( I/C + \hat{\mathbf{H}}^T\hat{\mathbf{H}} \right )^{-1}(\hat{\mathbf{H}}^T B\cdot \mathbf{T})  $
% \STATE \hspace{0.5cm}\textbf{return}  $\textbf{W},  \beta $
% \STATE 
% \STATE {\textsc{PREDICT}}$(\mathbf{X} )$
% \STATE \hspace{0.5cm}$ \mathbf{H} \gets  (\mathbf{X}^T\textbf{W} )/( \mathbb{1}\mathbf{X}  + \mathbb{1}\textbf{W}- \mathbf{X}^T\textbf{W}  ) $
% \STATE \hspace{0.5cm}\textbf{return}  $\textsc{sign}( \mathbf{H} \beta )$
% \end{algorithmic}
% \label{alg1}
% \end{algorithm}

% \section{Mathematical Typography \\ and Why It Matters}

\subsection{Ablation Study}

\begin{table*}[t]
\renewcommand\arraystretch{1.15}
  \caption{DERs (\%) of ablation experiments.}
  \footnotesize
  \label{tab:DERs1}
  \centering
\tabcolsep0.05in
\resizebox{\textwidth}{!}{
  \begin{tabular}{lcccccccccccc}
    \toprule
    \centering
    \multirow{3}*{Methods} & \multicolumn{6}{c|}{Simulated datasets} & \multicolumn{6}{c}{Real-word datasets} \\
    \cmidrule{2-7} \cmidrule{8-13}
    ~ & \multirow{2}*{\shortstack{Diarization \\ loss}} & \multicolumn{5}{c}{Number of speakers} & \multirow{2}*{\shortstack{Diarization \\ loss}} & CALLHOME & DIHARD II & DIHARD III & \multicolumn{2}{c}{AMI}\\ 
    \cmidrule{3-7} \cmidrule{9-9} \cmidrule{10-10} \cmidrule{11-11} \cmidrule{12-13}
    ~ & ~  & 1 & 2 & 3 & 4 & All ($\leq$ 4) & & $\leq$ 4 & $\leq$ 4 & $\leq$ 4 & Dev & Eval\\
    \midrule
% {LS-EEND} (prop.) & \multicolumn{1}{c}{{0.79}} & \multicolumn{1}{c}{{3.76}} & \multicolumn{1}{c}{\textbf{8.10}} & \multicolumn{1}{c}{\textbf{11.45}}\\
\multirow{3}*{\raisebox{-1.7ex} {LS-EEND (Sim\{1-4\}spk)}}  & \multirow{2}*{BCE} & \multirow{2}*{0.45} & \multirow{2}*{4.14} & \multirow{2}*{7.99} & \multirow{2}*{11.34} & \multirow{2}*{7.59} & PIT & \multicolumn{1}{c}{11.67} & 21.56 & 14.56  & 23.63 & 22.39\\
&  & & & & & & BCE & 14.07 & 24.48 & 16.45 & 40.53 & 36.08\\
\cmidrule{2-13}
   & PIT & \multicolumn{1}{c}{0.77} & \multicolumn{1}{c}{6.89} & \multicolumn{1}{c}{13.40} & \multicolumn{1}{c}{18.79} & 12.63 & PIT & 14.32 & 25.38 & 16.98  & 33.74 & 29.07\\
\midrule
% \quad  with appearance order & \multicolumn{1}{c}{{1.39}} & \multicolumn{1}{c}{{5.64}} & \multicolumn{1}{c}{12.74}  & \multicolumn{1}{c}{{19.21}} \\
\quad evaluation with appearance order & BCE & \multicolumn{1}{c}{0.45} & \multicolumn{1}{c}{{5.68}} & \multicolumn{1}{c}{11.39}  & \multicolumn{1}{c}{{17.46}} & 11.25 & - & - & - & - & -  & - \\
% \quad  w/o L2-normalization  & \multicolumn{1}{c}{\textbf{0.75}} & \multicolumn{1}{c}{3.83} & \multicolumn{1}{c}{8.56} & \multicolumn{1}{c}{11.88} \\ 
\quad  w/o L2-normalization & BCE  & \multicolumn{1}{c}{0.45} & \multicolumn{1}{c}{{4.10}} & \multicolumn{1}{c}{8.17} & \multicolumn{1}{c}{11.86} & 7.65 & PIT & 11.71 & 21.73 & 14.62  & 23.24  & 23.59 \\ 
 % \quad  w/o embedding similarity loss  & \multicolumn{1}{c}{1.54} & \multicolumn{1}{c}{4.57} & \multicolumn{1}{c}{10.35} & \multicolumn{1}{c}{14.50}\\
 \quad  w/o embedding similarity loss & BCE & \multicolumn{1}{c}{0.45} & \multicolumn{1}{c}{4.74} & \multicolumn{1}{c}{10.69} & \multicolumn{1}{c}{15.16} & 9.97 & PIT & 12.77 & 23.89  & 16.70 & 27.54 & 27.57\\
 % \quad  w/o look-ahead  & \multicolumn{1}{c}{1.27} & \multicolumn{1}{c}{4.95} & \multicolumn{1}{c}{10.28} & \multicolumn{1}{c}{14.35}\\
 \quad  w/o look-ahead & BCE & \multicolumn{1}{c}{0.49} & \multicolumn{1}{c}{5.48} & \multicolumn{1}{c}{10.67} & \multicolumn{1}{c}{14.03} & 9.68 & PIT & 12.76 & 22.97 & 15.74 & 29.07  & 27.71\\
 % \quad  w/o convolution module & \multicolumn{1}{c}{0.71} & \multicolumn{1}{c}{4.38} & \multicolumn{1}{c}{10.24} & \multicolumn{1}{c}{15.31}\\
  \quad  w/o convolution module & BCE & \multicolumn{1}{c}{0.46} & \multicolumn{1}{c}{4.94} & \multicolumn{1}{c}{11.87} & \multicolumn{1}{c}{16.77} & 10.97 & PIT & 13.26 & 22.91 & 16.42 & 28.73 & 29.32\\
 % \quad  with $\gamma$=Eq.(\ref{eq:gama})  & \multicolumn{1}{c}{0.96} & \multicolumn{1}{c}{\textbf{3.66}} & \multicolumn{1}{c}{8.86} & \multicolumn{1}{c}{18.30}\\  
  
  {\quad w/o non-speech marker} & BCE & 0.47 & 3.90 & 8.63 & 11.86 & 7.93 & PIT & 12.15 & 21.80 & 15.23 & 23.60 & 23.39 \\
  
  {\quad w/o termination marker} & BCE & 0.84 & 10.24 & 14.02 & 11.89 & 10.81 & PIT & 12.65 & 23.17 & 15.39 & 25.43  & 24.53\\

  \quad  with $\gamma$=Eq.(\ref{eq:gama}) & BCE & \multicolumn{1}{c}{0.37} & \multicolumn{1}{c}{3.92} & \multicolumn{1}{c}{8.95} & \multicolumn{1}{c}{18.74} & 10.68 & PIT & 17.61 & 29.57 & 23.88 & 41.63  & 45.37\\

  {\quad with masked self-attention} & BCE & \multicolumn{1}{c}{ 0.53} & \multicolumn{1}{c}{ 3.76} & \multicolumn{1}{c}{ 7.95} & \multicolumn{1}{c}{ 9.86} & 6.93 & PIT & 11.69 & 21.02 & 14.03 & 20.59 & 22.69 \\

  % \midrule
  % \color{blue}LS-EEND (Sim\{1-4\}spk) & \color{blue}1000 & \color{blue}PIT & \multicolumn{1}{c}{\color{blue}4.01} & \multicolumn{1}{c}{\color{blue}15.55} & \multicolumn{1}{c}{\color{blue}26.46} & \multicolumn{1}{c}{\color{blue}33.82} & - & - & - & - & - & - \\
  % \color{blue}LS-EEND (Sim\{1-4\}spk) & \color{blue}1000 & \color{blue}BCE & \multicolumn{1}{c}{\color{blue}5.23} & \multicolumn{1}{c}{\color{blue}19.40} & \multicolumn{1}{c}{\color{blue}33.34} & \multicolumn{1}{c}{\color{blue}42.51} & - & - & - & - & - & - \\
  % \color{blue}LS-EEND (Sim\{1-4\}spk) & \color{blue}10,000 & \color{blue}PIT & \multicolumn{1}{c}{\color{blue}1.80} & \multicolumn{1}{c}{\color{blue}9.07} & \multicolumn{1}{c}{\color{blue}18.10} & \multicolumn{1}{c}{\color{blue}25.73} & - & - & - & - & - & - \\
  % \color{blue}LS-EEND (Sim\{1-4\}spk) & \color{blue}10,000 & \color{blue}BCE & \multicolumn{1}{c}{\color{blue}1.50} & \multicolumn{1}{c}{\color{blue}7.38} & \multicolumn{1}{c}{\color{blue}16.04} & \multicolumn{1}{c}{\color{blue}23.64} & - & - & - & - & - & - \\
  % \color{blue}LS-EEND (Sim\{1-4\}spk) & \color{blue}100,000 & \color{blue}PIT & \multicolumn{1}{c}{\color{blue}0.77} & \multicolumn{1}{c}{\color{blue}6.89} & \multicolumn{1}{c}{\color{blue}13.40} & \multicolumn{1}{c}{\color{blue}18.79} & \color{blue}PIT & \color{blue}14.32 & \color{blue}25.38 &\color{blue}16.98  & \color{blue}33.74 & \color{blue}29.07\\
  
\bottomrule
\end{tabular}
}
\end{table*}

\begin{figure}[t]
\centering
\includegraphics[width=\linewidth]{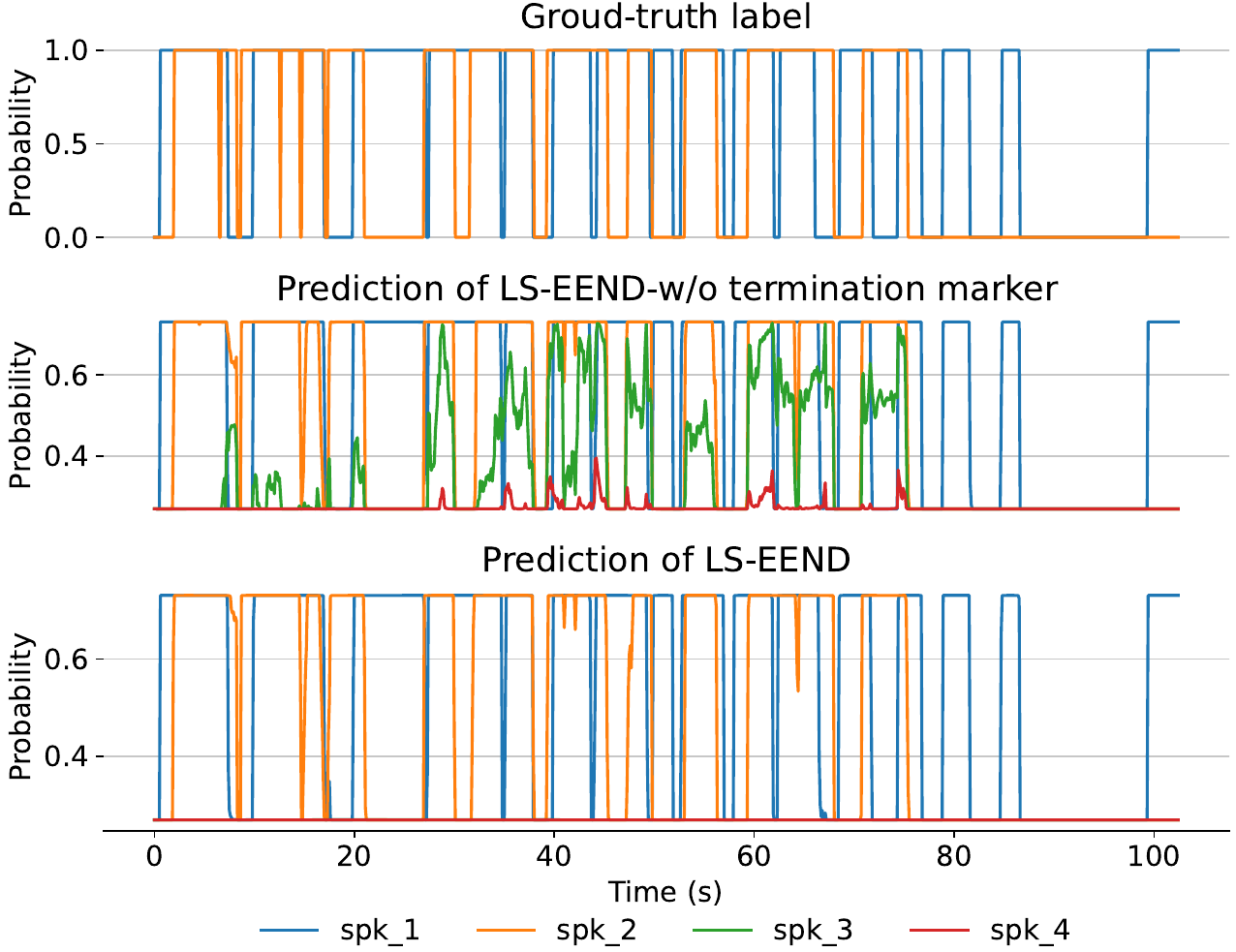}
\caption{An example of predicted speaker activities (for two ground truth speakers) when  using or not using the speaker termination marker.}
\label{fig:pred}
\end{figure}

Table \ref{tab:DERs1} shows the results of ablation experiments to verify the effectiveness of the various schemes proposed in this work. Due to the largely increased training cost with more speakers, all ablation studies are conducted on data with no more than 4 speakers. Specifically, the ablation models are pre-trained with up to Sim\{1-4\}spk data, and then adapted with real-world data containing no more than 4 speakers. 
% \addnote[real-world data]{1}{Each model is further fine-tuned on real-world data and evaluated under scenarios involving no more than four speakers.}

We first compare when training the model with BCE loss (according to the speaker appearance order) or PIT loss. Note that the diarization performance is always measured with the best matching speaker order. We can see that BCE noticeably outperforms PIT for pre-training with simulated datasets. However, when adapting the BCE pre-trained model to real-world datasets, PIT noticeably outperforms BCE. Due to the worse performance of PIT pre-training, PIT pre-training plus PIT adaption does not perform as well as BCE pre-training plus PIT adaption. Appearance order is a natural arrangement of speaker order for online diarization, which is also consistent with the causal learning of the diarization network, in the sense that the network can sequentially arrange speaker based on historical information for both speaker embedding learning and attractor learning. Note that, in EEND, the changing/local speaker embeddings are learned based on certain rules, which means the order of multiple speaker embeddings also does matter. However, generally speaking, BCE training is more challenging than PIT training, as BCE forces a certain constraint compared to PIT. When training with a large amount of simulated data, both BCE and PIT can be well trained to have a small generalization error from training to test. In addition, BCE achieves better speaker embeddings than PIT, as we have observed considerably smaller embedding similarity loss for BCE than for PIT (the embedding similarity losses are not shown here). As a result, BCE noticeably outperforms PIT on simulated test sets. By contrast, when adapting the model with a small amount of real-world data, the more challenging BCE task cannot be well trained, as it has a larger generalization error (which is not shown here) than PIT. Overall, using BCE with a large amount of simulated data for pre-training and using PIT for adaption (if only a small amount of real-world data are available) is the optimal training configuration for the proposed model.

% \addnote[bce pit]{1}{To analyze the applicability of BCE (according
% to the speaker appearance order) and PIT diarization losses to the LS-EEND model, we train the model using 1,000 (155 h), 10,000 (1,550 h), and 100,000 (15,500 h) simulated mixtures, respectively. The results presented in the lower part of Table \ref{tab:DERs1} show that BCE performs worse than PIT when meeting small-scale data. However, as the data size increases, BCE gradually outperforms PIT. When trained with a large amount of data (100,000 mixtures), BCE performs significantly better than PIT. We believe this is due to BCE with appearance order tackling a more challenging task. Specifically, it requires clustering embeddings belonging to the same speaker (ignoring overlapping speech for the sake of simplicity in description), in addition to assigning speaker IDs based on their order of appearance. Tackling such a more complex task typically demands a more sophisticated network architecture and larger datasets for effective learning. Once this challenging task is well-learned, the model tends to perform better on the simpler DER metric that relies on optimal matching between predictions and ground-truth labels. The DERs on real-world data are also provided, demonstrating that using BCE with a large amount of simulated data and PIT with a small amount of real-world data is the optimal diarization loss configuration.}

Next, on top of the LS-EEND (Sim\{1-4\}spk) model with the optimal training configuration, i.e. BCE pre-training plus PIT adaption, we conduct ablation experiments by either removing one component (w/o one component) or replacing one component with an alternative (with an alternative). When evaluated \emph{with appearance order}, the DERs get higher. This performance degradation reflects the enrollment failure rate for speakers appearing, with the failure rate rising as the number of speaker increases. The effectiveness of L2-normalization, embedding similarity loss, look-ahead, convolution module in Conformer, non-speech marker and termination marker are validated, as the DERs get higher (for most conditions) for both simulated and real-world data when any of these components is removed. 
When removing the embedding similarity loss, the performance degradation on simulated data becomes more prominent as the number of speakers increases, which indicates that it is indeed more difficult for distinguishing more speakers, and the proposed embedding similarity loss helps to mitigate this problem. The performance degradation is also significant for the longest AMI data, which indicates that the embedding similarity loss is also helpful for maintaining the consistence of speaker embeddings along a very long period.
The considerable performance degradation caused by removing the non-speech marker suggests that the non-speech attractor helps improve the discriminability of speech and non-speech frames to some extent.
Removing the termination marker causes a significant performance degradation when the number of active speakers is smaller than the maximum number (4 in this experiment), such as the results of 2 and 3 speakers in the simulated test sets. Fig. \ref{fig:pred} shows an example where fake speakers are detected when there is no termination marker. In contrast, the detection of active speakers can be correctly terminated when the termination marker is used.  
% removing it causes a significant performance degradation, which results from a high false alarm rate. This happens because when the actual number of active speakers is smaller than the maximum number of speakers in the training set, virtual active speakers are decoded, as shown in Fig. \ref{fig:pred}. The role of the termination marker is to determine the actual active speakers. For a maximum of four speakers, a total of six attractors are decoded. Excluding the probability associated with the non-speech attractor, five attractors are evaluated for speaker activity. In the case of two actual speakers, although five activity probabilities are produced, the network automatically predicts the third to fifth attractors as inactive.
 
% To not use of look-ahead, we modify the look-ahead convolution to be causal, with a kernel size of 10. 
To verify our new setting of $\gamma=1$ in Retention, its performance is compared with the original $\gamma$ setting as specified in (\ref{eq:gama}). The significant performance degradation on simulated data with 4 speakers and on real-world data when setting $\gamma$=Eq.(\ref{eq:gama}) indicates that information decay over time is detrimental to long-speech diarization. When we replace the Retention module with masked self-attention, performance can be slightly improved, but with a quadratic temporal complexity. Note that the audio length when fine-tuning on AMI is increased only up to 800 s, due to the limit of GPU memory.

\begin{figure}[t]
\centering
\includegraphics[width=0.9\linewidth]{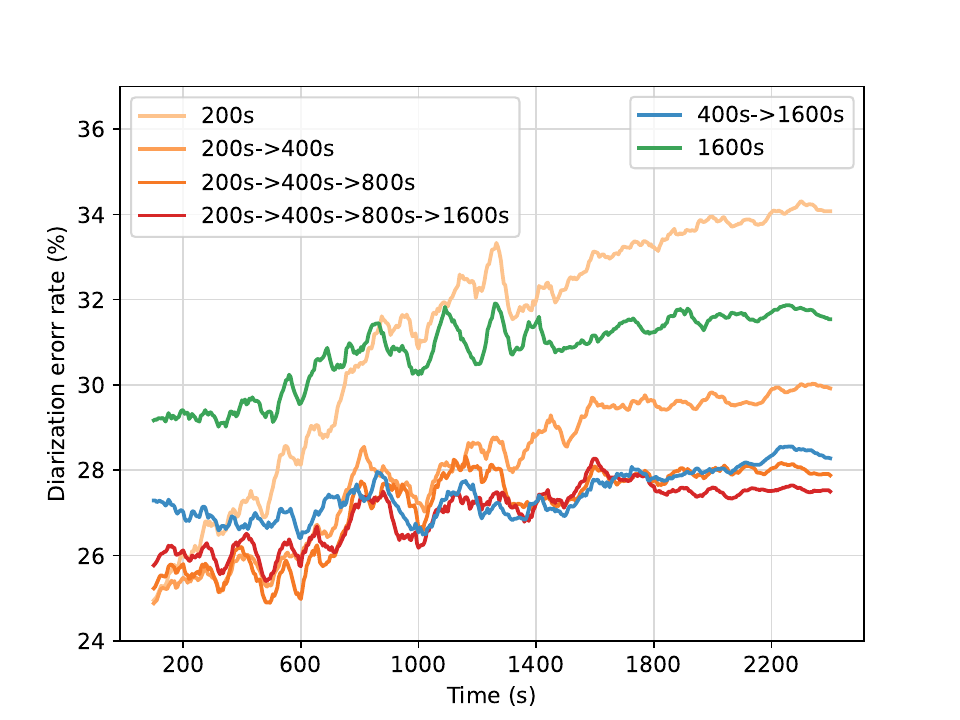}
\caption{DER curve with respect to audio length of LS-EEND (Sim\{1-4\}spk)-w/o embedding similarity loss, for the AMI Dev set. `-$>$' denotes the next step of adaptation.}
\label{fig:wo_emb_loss_der}
\end{figure}

In Fig. \ref{fig:wo_emb_loss_der},  we evaluate the effectiveness of the proposed progressive training strategy on one of the baseline systems, LS-EEND(Sim\{1-4\})-w/o embedding similarity loss. It is demonstrated that the proposed training strategy remains effective when applied to alternative models.

% \begin{figure}[t]
%     \centering
    
%     % 第一行
%     \begin{subfigure}{0.49\linewidth}
%         \centering
%         \includegraphics[width=\linewidth]{figs/ls_eend_4spk_der_dev_smth.pdf}
%         \captionsetup{labelformat=empty} % 取消标号
%         \caption{(a) proposed model} % 使用英文替代中文标题
%         \label{fig:subfig31}
%     \end{subfigure}
%     \hspace*{-0.1cm}
%     \begin{subfigure}{0.49\linewidth}
%         \centering
%         \includegraphics[width=\linewidth]{figs/wo_emb_loss_der_dev_smth.pdf}
%         \captionsetup{labelformat=empty} % 取消标号
%         \caption{(b) w/o L2-norm} % 使用英文替代中文标题
%         \label{fig:subfig32}
%     \end{subfigure}
%     % \hspace*{-0.1cm}

%     \caption{t-SNE visualization \cite{van2008visualizing} of speaker embeddings in 2-dimensional space, for the case of 4 speakers. Scatter points with grey borders and multiple colors represent overlapping frames.
%     % (a) proposed model, (b) w/o L2-normalization, (c) w/o embedding similarity loss, (d) w/o look-ahead, (e) w/o convolution module,  (f) $\gamma$=Eq.(\ref{eq:gama}). 
%     } % 整个图的英文标题
%     \label{fig:abl}
% \end{figure}

\begin{figure}[t]
    \centering
    
    % 第一行
    \begin{subfigure}{0.49\linewidth}
        \centering
        \includegraphics[width=\linewidth]{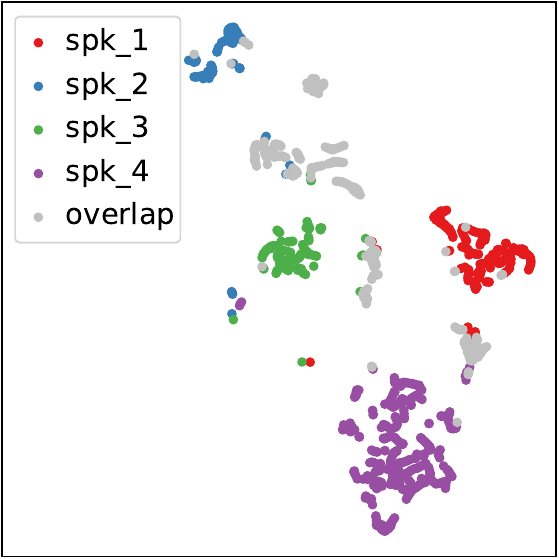}
        \captionsetup{labelformat=empty} % 取消标号
        \caption{(a) proposed model} % 使用英文替代中文标题
        \label{fig:subfig31}
    \end{subfigure}
    \hspace*{-0.1cm}
    \begin{subfigure}{0.49\linewidth}
        \centering
        \includegraphics[width=\linewidth]{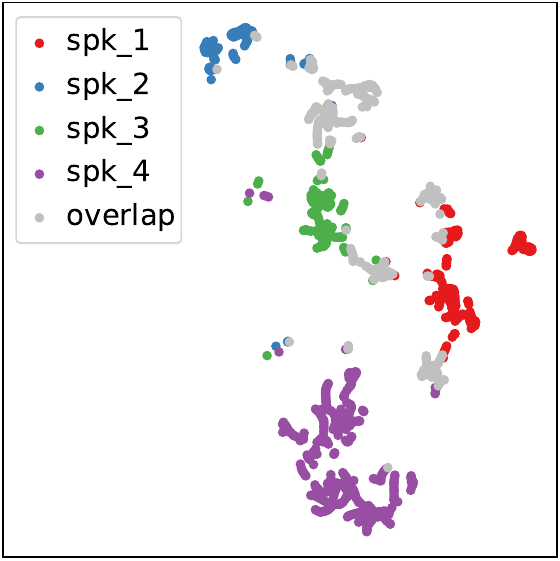}
        \captionsetup{labelformat=empty} % 取消标号
        \caption{(b) PIT (pre-)training} % 使用英文替代中文标题
        \label{fig:subfig32}
    \end{subfigure}
    % \hspace*{-0.1cm}
    \vspace{0.02cm} % 行间距，可以根据需要调整
    
    % 第二行
    \begin{subfigure}{0.49\linewidth}
        \centering
        \includegraphics[width=\linewidth]{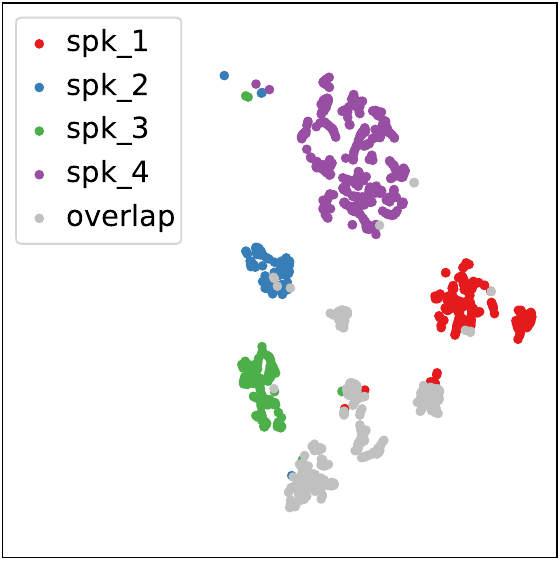}
        \captionsetup{labelformat=empty} % 取消标号
        \caption{(c) w/o similarity loss } % 使用英文替代中文标题
        \label{fig:subfig33}
    \end{subfigure}
    \hspace*{-0.1cm}
    \begin{subfigure}{0.49\linewidth}
        \centering
        \includegraphics[width=\linewidth]{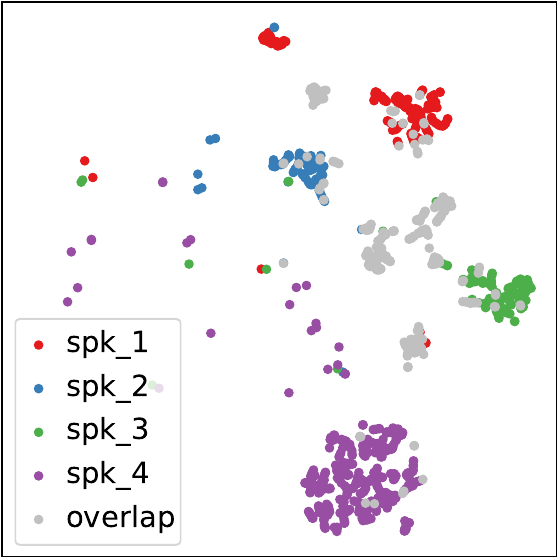}
        \captionsetup{labelformat=empty} % 取消标号
        \caption{(d) w/o look-ahead} % 使用英文替代中文标题
        \label{fig:subfig34}
    \end{subfigure}
    % \hfill
    \vspace{0.02cm} % 行间距，可以根据需要调整

    % 第三行
    % \hspace*{-0.1cm}
    \begin{subfigure}{0.49\linewidth}
        \centering
        \includegraphics[width=\linewidth]{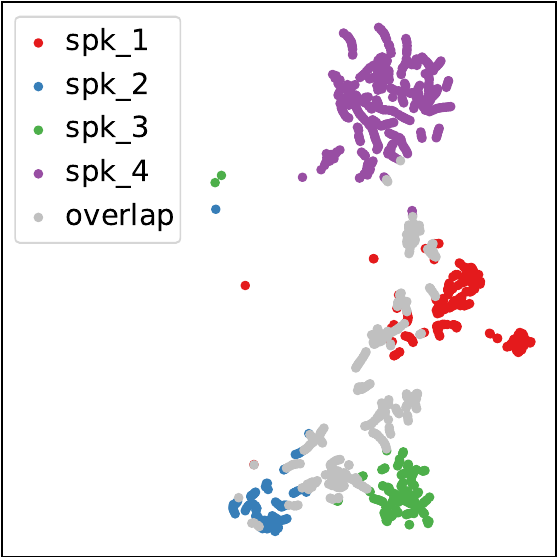}
        \captionsetup{labelformat=empty} % 取消标号
        \caption{(e) w/o convolution} % 使用英文替代中文标题
        \label{fig:subfig35}
    \end{subfigure}
    \hspace*{-0.1cm}
    \begin{subfigure}{0.49\linewidth}
        \centering
        \includegraphics[width=\linewidth]{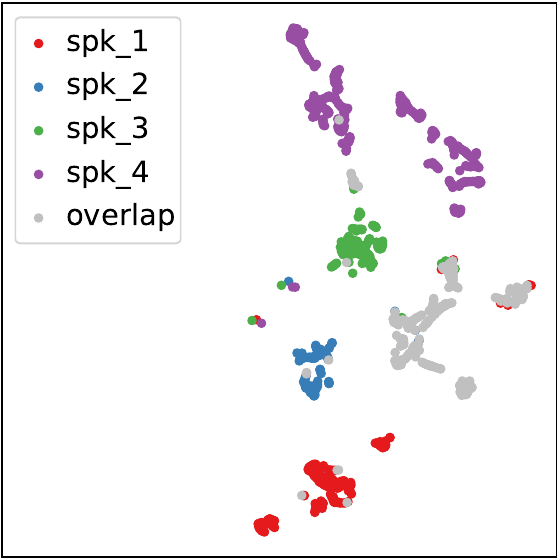}
        \captionsetup{labelformat=empty} % 取消标号
        \caption{(f) $\gamma$=Eq.(\ref{eq:gama})} % 使用英文替代中文标题
        \label{fig:subfig36}
    \end{subfigure}

    \caption{t-SNE visualization \cite{van2008visualizing} of speaker embeddings in 2-dimensional space, for the case of 4 speakers. 
    % Scatter points with grey borders and multiple colors represent overlapping frames.
    % (a) proposed model, (b) w/o L2-normalization, (c) w/o embedding similarity loss, (d) w/o look-ahead, (e) w/o convolution module,  (f) $\gamma$=Eq.(\ref{eq:gama}). 
    } % 整个图的英文标题
    \label{fig:abl}
\end{figure}

Furthermore, Fig. \ref{fig:abl} visualizes and compares the speaker embeddings for different ablation cases. 
% \addnote[overlap]{1}{Each color denotes a unique speaker. Overlapping frames are represented by scatter points with grey borders and multiple colors, indicating the set of speakers active during those frames.} 
It can be seen that each component enhances the embedding distribution with better clustering of same speakers and discrimination between different speakers. Specifically, PIT (pre-)training achieves less concentrated intra-speaker embeddings and less distinct embeddings between single-speaker and overlapped speakers. The embedding similarity loss  improves clustering for individual speaker and also causes embeddings of overlapping speech to be distributed in between single-speaker embeddings. The look-ahead mechanism helps to remove some outliers. When $\gamma$=Eq.(\ref{eq:gama}), the embeddings for same speakers exhibit noticeable discontinuities, indicating that information decay can impair the model's ability to maintain speaker consistency over long periods of time. 
% Therefore, $\gamma$ is set to 1 to retain the full left context.

\section{Conclusion}
This paper proposes the LS-EEND model, for long-form streaming end-to-end neural diarization. The key point of LS-EEND is to model speakers within the self-attention-based decoder along both the time and speaker dimensions, which enables information retrieval from historical frames and in between speakers. Another important point of LS-EEND is the adaptation of Retention, which ensures the exploitation of very long-term dependencies for the same speaker and meanwhile the constant computation cost. 
With the multi-step progressive training strategy, the proposed model is able to handle a high number of speakers and very long audio recordings, which is crucial for real applications. Note that the maximum number of speakers is arbitrarily set to 8 in this work. As shown in the present experiments that diarization performance is reasonably decreased when the number of speakers is increased to 8, we believe that it can be further increased if necessary, with a reasonable performance decrease.

\bibliographystyle{IEEEtran}
\bibliography{mybib}

\newpage

 

% \end{IEEEbiography}



\vfill

\end{document}